\documentclass[preprint,showpacs,titlepage,aps,prd,
tightenlines,amsmath,byrevtex,nofootinbib]{revtex4}

\usepackage{graphicx}

\def\lsim{\raise0.3ex\hbox{$\;<$\kern-0.75em\raise-1.1ex
\hbox{$\sim\;$}}}
\def\gsim{\raise0.3ex\hbox{$\;>$\kern-0.75em\raise-1.1ex
\hbox{$\sim\;$}}}

\begin{document}

\preprint{hep-ph/0701070}

\title{Neutrinos; Opportunities and Strategies in the Future\protect\footnote{
Combined and slightly expanded written version of the invited talks given at 
International Conference ``Heavy Quarks and Leptons'', Munich, Germany, 
October 16-20, 2006, and 
Second World Summit on ``Physics Beyond the Standard Model'', 
Galapagos Islands, Ecuador, June 22-25, 2006 (post-deadline version).
}}

\author{Hisakazu Minakata}
\email{minakata@phys.metro-u.ac.jp}
\affiliation{
Department of Physics, Tokyo Metropolitan University, Hachioji, Tokyo 192-0397, Japan
}

\date{January 10, 2007}

\vglue 1.4cm

\begin{abstract}

I try to give an overview of future prospects of the experimental 
exploration of the unknowns in the neutrino mass pattern and the 
lepton flavor mixing. 
Because of the nature of the lectures on which this manuscript is based, 
I give some pedagogical discussions to prepare for the presentation 
in the later part. I start from measuring $\theta_{13}$ by reactors 
and accelerators as a prerequisite for proceeding to search for leptonic 
CP violation. I then discuss how CP violation can be uncovered, 
and how the neutrino mass hierarchy can be determined. 
I do these by resolving so called the ``parameter degeneracy'' 
which is necessary anyway if one wants to seek precision measurement 
of the lepton mixing parameters. 
As a concrete setting for resolving the degeneracy I introduce the 
Tokai-to-Kamioka-Korea two detector complex which receives 
neutrino superbeam from J-PARC, sometimes dubbed as ``T2KK''.
It is shown that T2KK is able to resolve all the eight-fold parameter 
degeneracy in a wide range of the lepton mixing parameters. 
I also discuss an alternative way for lifting the $\theta_{23}$ octant 
degeneracy by reactor-accelerator combined method. 
Finally, I discuss by taking some examples how some theoretically 
or phenomenologically motivated ideas can be tested experimentally.

\end{abstract}


\maketitle


\section{Introduction}

Neutrino physics, in particular its experimental part, has been 
extremely successful in the last 10 years. 
It would be worthwhile to look it back on this occasion
as a prologue to our discussion on the future. 
In 1998 the Super-Kamiokande atmospheric neutrino observation confirmed 
\cite{SKatm}  the smoking gun evidence for atmospheric neutrino anomaly 
seen in the deficit of the rate and in the zenith angle distribution of 
$\nu_{\mu}$ induced events in the Kamiokande experiment \cite{Kam-atm}.  
It was the first evidence for mass-induced neutrino oscillation.\footnote{
The history of theory of neutrino oscillation is somewhat involved.
In 1957 Pontecorvo \cite{Pontecorvo} discussed 
$\nu \leftrightarrow \bar{\nu}$ oscillation in close analogy to 
$K^0 \leftrightarrow \bar{K}^0 $ oscillation \cite{Gell-Mann}.  
In 1962
Maki, Nakagawa, and Sakata \cite{MNS} 
first pointed out the possibility of neutrino flavor transformation, 
the phenomenon established experimentally only recently 
as described here.  
}
The evidence for neutrino oscillation was readily confirmed by 
the first long-baseline (LBL) accelerator neutrino experiment K2K 
\cite{K2K} using man-made $\nu_{\mu}$ beam.
In this sense, the first corner stone was placed in the 2-3 sector of the 
lepton flavor mixing matrix, the MNS matrix \cite{MNS}. 
It established surprisingly that the mixing angle $\theta_{23}$ is large, 
which may even close to the maximal, refuting the prejudice of 
small flavor mixing angles deeply rooted among theorists at that time.

On the other hand, there have been great amount of efforts in 
the solar neutrino observation pioneered by Davis with his 
$^{37}$Cl  experiment in sixties which was developed 
in close collaboration with the devoted theorist \cite{Bahcall-Davis}. 
In the last 20 years the field has been enriched by participation 
by Kamiokande, Ga, Super-Kamiokande, and SNO experiments 
\cite{solar}. 
In particular, the latter two experiments were united to the confirm 
the particle physics nature of the solar neutrino problem, 
the evidence for solar neutrino flavor transformation \cite{SK-SNO}. 
Later SNO {\em in situ} confirmed the evidence \cite{SNO-direct}. 
I would like to note here that the deficit of the $^8$B flux obtained 
by Davis in his $^{37}$Cl experiment \cite{Cl}, 
though suffered from stubborn skepticism for more than 30 years, 
was convincingly confirmed by the SNO charged current (CC) measurement. 
The beautiful finale of the solar neutrino problem came with 
KamLAND \cite{KamLAND} which identified its cause as due to 
the mass-induced neutrino oscillation which clearly pinned down 
the large mixing angle (LMA) MSW solar neutrino solution \cite{MSW}. 
The particular significance of the KamLAND result in this context, 
so called the KamLAND massacre (of non-standard scenarios), 
was emphasized by many people with detailed analysis for example in \cite{valle}. 
The resultant mixing angle $\theta_{12}$ turned out to be 
large, but not maximal. 

Finally, several experiments, 
Super-Kamiokande \cite{atm_evidence}, KamLAND \cite{KL_evidence}, 
K2K \cite{K2K_evidence}, and MINOS \cite{MINOS}, 
observed the oscillatory behavior, thereby established the 
phenomenon of mass-induced neutrino oscillation. 
At this moment, it constitutes the first and the unique evidence 
for physics beyond the Standard Model.

What is the next?
The most common answer, which I also share, is to explore 
the unknown 1-3 sector, for which the only knowledge we have 
is the upper bound on $\theta_{13}$ \cite{CHOOZ,K2K_app}. 
Discovery of leptonic CP violation must throw light on tantalizing 
mystery of interrelationship between quarks and leptons. 
The discussion of the quark-lepton correspondence which can 
be traced to early sixties \cite{QLcorr}, and in a modern context 
presented in a compelling form with the anomaly cancellation in 
Standard Model \cite{anomcancel} strongly suggests that they 
have common roots. 
It is also possible that the Kobayashi-Maskawa type CP violation \cite{KM} 
in the lepton sector might be related to CP violation at high energies 
which is required for  leptogenesis \cite{leptogenesis} to work. 
See, e.g., \cite{petcov} and the references cited therein for this point.

Despite the great progress in our understanding mentioned above 
we do have many important unanswered questions. 
The list includes, for example, the followings: 
What is the origin of neutrino masses and mixing?
What is the reason for disparity between small quark and large lepton mixings?  
Is there underlying quark-lepton symmetry, or quark-lepton complementarity?
Is there flavor symmetry which includes quarks and leptons? 
I am sure that many more questions exist. 
These points are discussed in depth in a recent review \cite{moha-smi}.

The new stage of neutrino physics may also be characterized as 
beginning of the era of precision measurement of lepton mixing parameters. 
Testing the various theoretical ideas proposed to understand the 
uncovered structure mentioned above requires accurate 
determination of mixing parameters. 
For example, to test the quark-lepton complementarity \cite{QLC,QLC_rev} 
experimentally, 
one needs to improve accuracies for $\theta_{12}$ determination 
from the current one, $\simeq 12$\% for $\sin^2 \theta_{12}$, to 
the one comparable to the Cabibbo angle, $\sim 1$\% \cite{PDG}. 
It will be discussed in Sec.~\ref{proceed}.

I must admit that the scope of my discussions is quite limited; 
The crucially important issues such as absolute neutrino mass, 
nature of neutrinos (Majorana vs. Dirac), Majorana CP violation and 
leptogenesis are not covered. Moreover, it covers only a part of the 
things that should be addressed for exploring unknowns done by 
the future LBL experiments, that is,  
concrete ways of how to determine the mixing parameters with the 
next generation conventional $\nu_{\mu}$ superbeam \cite{lowECP,superbeam2} 
and reactor experiments.  
Yet, conventional superbeam experiments are extremely interesting 
because, in principle, they can be done in the next 10-15 years 
without long-term R$\&$D effort.

Here is a composition of this long report. 
First of all, I intend to be pedagogical in writing this report; 
I met with many brilliant young people in ``World Summit in Galapagos'' 
\cite{galapagos}, 
and a broad class of audiences who keenly interested in neutrino 
physics in ``Heavy Quarks and Leptons'' \cite{HQL}, 
and it is a pity if this manuscript is entirely unreadable to them.
In Sec.~\ref{theta23}, we review how the atmospheric parameters 
$\Delta m^2_{32}$ and $\theta_{23}$ are determined. 
In Sec.~\ref{theta13}, we explain how $\theta_{13}$ can be measured 
and briefly review the reactor and the accelerator methods. 
In Sec.~\ref{vac-matter}, we provide a simple understanding 
of the interplay between the vacuum and the matter effect by 
introducing the bi-probability plot. 
In Sec.~\ref{delta}, we mention two alternative strategies of how to 
measure CP violation and give some historical remarks on how the 
thoughts on measuring leptonic CP violation were evolved. 
In Sec.~\ref{Pdege}, we explain in a simple terms the cause of 
the parameter degeneracy by using the bi-probability plot.
It is an important topics for precision measurement of the lepton mixing 
parameters because the degeneracy acts as a notorious obstacle to it. 
In Sec.~\ref{resolveD}, we discuss how the eight-fold parameter degeneracy 
can be resolved {\em in situ} by using ``T2KK'', the Tokai-to-Kamioka-Korea setting. 
In Sec.~\ref{reactor-accelerator}, we describe an alternative method 
for solving a part of the degeneracy called the $\theta_{23}$ octant 
degeneracy by combining reactor and accelerator experiments. 
In Sec.~\ref{proceed}, we discuss, by taking a concrete example, 
how theoretical/phenomenological 
hypothesis can be confronted to experiments. 
In Sec.~\ref{conclude}, we give a concluding remark.

\section{Atmospheric parameters; $\Delta m^2_{31}$ and $\theta_{23}$}
\label{theta23}

``Bread and butter'' in the coming era of precision measurement 
of lepton mixing parameters is the accurate determination of the 
atmospheric parameters, $\Delta m^2_{31}$ and $\theta_{23}$. 
It will be carried out by the accelerator disappearance experiments 
which measures energy spectrum modulation of muon neutrinos. 
Ignoring terms proportional to $\Delta m^2_{21}$ and $\theta_{13}$, 
the disappearance probability in vacuum can be written as 
\begin{eqnarray}
P(\nu_{\mu} \rightarrow \nu_{\mu}) & = & 1 - 
\sin^2 2\theta_{23}\sin^2\Bigl(\frac{\Delta m^2_{31} L}{4E} \Bigr)
\label{Pvac}
\end{eqnarray}
In view of (\ref{Pvac}), 
very roughly speaking, the position and the depth of the dip 
corresponding to the first oscillation maximum, 
$\Delta_{31} \equiv \frac{\Delta m^2_{31} L}{4 E} = \pi/2$, 
tell us $\Delta m^2_{31}$ and $\sin^2 2\theta_{23}$, respectively.

The current limits on these parameters from the SK atmospheric 
neutrino observation are 
$1.5 \times 10^{-3} \text{eV}^2 < \Delta m^2_{31} < 3.4 \times 10^{-3} \text{eV}^2$ 
and $\sin^2 2\theta_{23} > 0.92$ at 90\% CL \cite{SKatm}. 
K2K, the first accelerator LBL experiment obtained the similar 
results, 
$1.9 \times 10^{-3} \text{eV}^2 < \Delta m^2_{31} < 3.5 \times 10^{-3} \text{eV}^2$ 
at 90\% CL 
though the sensitivity to $\theta_{23}$ is much worse, 
$\sin^2 2\theta_{23} < 0.6$  \cite{K2K_final}. 
The currently running MINOS experiment \cite{MINOS} aims at 
determining $\Delta m^2_{31}$ to $\simeq6$\% level, and 
$\sin^2 2\theta_{23}$ to $\simeq8$\% level, both at 90\% CL. 
The next generation LBL experiment T2K \cite{T2K} is expected 
to improve the sensitivity to 
$\simeq2$\% level for $\Delta m^2_{31}$ excluding systematics, 
and 
$\simeq1$\% level for $\sin^2 2\theta_{23}$ including systematics, 
both at 90\% CL \cite{hiraide}. 
These numbers are cross checked in various occasions \cite{MSS04,resolve23}. 
The US project NO$\nu$A \cite{NOVA} will also have the similar sensitivities. 
These accuracies are quite essential to resolve the parameter degeneracy 
(see Sec.~\ref{Pdege}) to achieve the goal of precision determination 
of the lepton mixing parameters.

\section{$\theta_{13}$}
\label{theta13}

To reach the goal of seeing leptonic CP violation, we have to clear 
the first hurdle, knowing the value of $\theta_{13}$. 
What is the most appropriate way to measure the parameter? 
To answer the question we consider the neutrino oscillation channel 
which involve $\nu_{e}$, otherwise $\theta_{13}$ would not be 
contained in leading order. 
There are two candidate channels; 
$\nu_{e} \rightarrow \nu_{e}$ and 
$\nu_{\mu} \rightarrow \nu_{e}$ 
(or, $\nu_{e} \rightarrow \nu_{\mu}$). 
In our discussion that follows, $\nu_{e} \rightarrow \nu_{\tau}$ is 
the same as $\nu_{e} \rightarrow \nu_{\mu}$.

We note that $P(\nu_{e} \rightarrow \nu_{e})$ probed at energy/baseline 
appropriate to atmospheric $\Delta m^2$ scale
consists of interference terms between amplitudes 
$A(\nu_{e} - \nu_{3} \rightarrow  \nu_{3} - \nu_{e})$ and 
$A(\nu_{e} - \nu_{1} \rightarrow  \nu_{1} - \nu_{e}) + 
A(\nu_{e} - \nu_{2} \rightarrow  \nu_{2} - \nu_{e})$. 
Then, obviously $|U_{e3}|^2$ is involved in the disappearance probability. 
On the other hand, in the appearance channel 
$\nu_{\mu} \rightarrow \nu_{e}$, the oscillation probability contains 
interference terms between amplitudes 
$A(\nu_{\mu} - \nu_{3} \rightarrow  \nu_{3} - \nu_{e})$ and 
$A(\nu_{\mu} - \nu_{1} \rightarrow  \nu_{1} - \nu_{e}) + 
A(\nu_{\mu} - \nu_{2} \rightarrow  \nu_{2} - \nu_{e})$. 
Then, the appearance channel looks to be advantageous because 
only a single power of $|U_{e3}|$ is involved. 
But, it is untrue; When there is a hierarchy in $\Delta m^2$,  
$\Delta m^2_{21} \ll \Delta m^2_{31}$, 
unitarity tells us that these two terms nearly cancel, leaving 
another power of $|U_{e3}|$. As a consequence, 
$P(\nu_{\mu} \rightarrow \nu_{e})$ is also proportional to $|U_{e3}|^2$. 
Hence, there are two comparably good ways to measure $\theta_{13}$; 
the reactor and the accelerator methods which measure 
$P(\nu_{e} \rightarrow \nu_{e})$ and $P(\nu_{\mu} \rightarrow \nu_{e})$, 
respectively.
Let us describe them one by one.

Before getting into the task we give here the explicit expressions 
of oscillation probabilities. For $\bar{\nu}_{e}$ disappearance channel 
it reads (see e.g., erratum in \cite{MSYIS}) 
\begin{eqnarray}
1-P(\bar{\nu}_{e} \rightarrow \bar{\nu}_{e}) &=&
\sin^2{2 \theta_{13}}
\sin^2 \left( {\Delta m^2_{31}L \over 4E} \right) 
- \frac{1}{\,2\,} s^2_{12}
\sin^2{2 \theta_{13}}
\sin{\left( {\Delta m^2_{31}L \over 2 E} \right) }
\sin{\left( {\Delta m^2_{21}L \over 2 E} \right) }
\nonumber \\
&+&
\left[
c^4_{13} \sin^2{2 \theta_{12}} +
s^2_{12} \sin^2{2 \theta_{13} \cos{\left( {\Delta m^2_{31}L \over 2 E} \right) }}
\right]
\sin^2 \left( {\Delta m^2_{21}L \over 4E} \right). 
\label{Pvac_ee}
\end{eqnarray}
For the appearance channel, we use the 
$\nu_{\mu} (\bar{\nu}_\mu) \to \nu_{e} (\bar\nu_e$) 
oscillation probability with first-order matter effect \cite{AKS}
\begin{eqnarray}
P[\nu_{\mu}(\bar{\nu}_{\mu}) \rightarrow 
\nu_{\rm e}(\bar{\nu}_e)] &=&  
%
\sin^2{2\theta_{13}} s^2_{23}
\left[
\sin^2 \left(\frac{\Delta m^2_{31} L}{4 E}\right)
-\frac {1}{2}
s^2_{12}
\left(\frac{\Delta m^2_{21} L}{2 E}\right)
\sin \left(\frac{\Delta m^2_{31} L}{2 E}\right) 
\right.
\nonumber \\
&&\hspace*{24mm} {}\pm
\left.
\left(\frac {4 Ea}{\Delta m^2_{31}}\right)
\sin^2 {\left(\frac{\Delta m^2_{31} L}{4 E}\right)}
\mp 
\frac{aL}{2}
\sin \left(\frac{\Delta m^2_{31} L}{2 E}\right) 
\right]
\nonumber \\
&+& 
2J_{r} \left(\frac{\Delta m^2_{21} L}{2 E} \right)
\left[
\cos{\delta}
\sin \left(\frac{\Delta m^2_{31} L}{2 E}\right) \mp 
2 \sin{\delta}
\sin^2 \left(\frac{\Delta m^2_{31} L}{4 E}\right) 
\right] 
\nonumber \\
&+&
c^2_{23} \sin^2{2\theta_{12}} 
\left(\frac{\Delta m^2_{21} L}{4 E}\right)^2,
\label{Pmue}
\end{eqnarray}
where the terms of order 
$s_{13} \left( \frac{\Delta m^2_{21}}{\Delta m^2_{31}} \right)^2$ 
and 
$aL s_{13} \left( \frac{\Delta m^2_{21}}{\Delta m^2_{31}} \right)$ 
are neglected. 
In Eq.~(\ref{Pmue}), $a\equiv \sqrt 2 G_F N_e$  \cite{MSW}
where $G_F$ is the Fermi constant, $N_e$ denotes the averaged 
electron number density along the neutrino trajectory in 
the earth, 
$J_r$ $(\equiv c_{12} s_{12} c_{13}^2 s_{13} c_{23} s_{23} )$ 
denotes the reduced Jarlskog factor, and the upper and the 
lower sign $\pm$ refer to the neutrino and 
anti-neutrino channels, respectively. 
%
In both of the oscillation probabilities, 
$P(\bar{\nu}_{e} \rightarrow \bar{\nu}_{e})$ and 
$P(\nu_{\mu} \rightarrow \nu_{e})$, 
the leading atmospheric oscillation terms have the common factor 
$\sin^2{2\theta_{13}}$, in agreement with the discussion given above. 
The last term in Eq.~(\ref{Pmue}) is the solar scale oscillation term, 
which will be important for resolving the $\theta_{23}$ degeneracy.

\subsection{Reactor measurement of $\theta_{13}$}

It was proposed \cite{krasnoyarsk,MSYIS} 
that by using identical near and far detectors which 
is placed close to and at around $\sim$1 km from the reactor, 
respectively, one can search for non-zero $\theta_{13}$ to a region of 
$\sin ^2 2\theta_{13} \sim 0.01$. 
An advantage of the reactor $\theta_{13}$ experiments is their 
cost effectiveness which stems from that the beam is intense enough 
(and furthermore free!) and low in energy to allows relatively compact detectors 
placed at baselines much shorter than those of accelerator experiments.
Intensive efforts over several years from these proposals 
entailed the various projects in world wide as described in 
\cite{reactor_white}. 
By now a few projects have already been approved, or are close 
to the status \cite{DCHOOZ,RENA,Daya-Bay}.

Scientific merit of the reactor measurement of $\theta_{13}$ is 
that it provides pure measurement of $\theta_{13}$ without 
being affected by other mixing parameters, as emphasized in 
\cite{MSYIS}. 
It implies, among other things, that it can help resolving 
the $\theta_{23}$ octant degeneracy as pointed out in \cite{MSYIS}, 
and recently demonstrated in detail in \cite{resolve23}. 
On the other hand, the same property may be understood as 
``shortcoming'' of the reactor experiment, if one want to search for 
leptonic CP violation. 
It is known that $\nu_{e}$ ($\bar{\nu}_{e}$) disappearance 
probability has no sensitivity to $\delta$ even in matter with arbitrary profile 
with negligible higher order correction \cite{solarCP}. 
We note, however, that reactor $\theta_{13}$ experiment 
can be combined with accelerator appearance measurement 
to uncover CP violation \cite{reactorCP}.

\subsection{Accelerator measurement of $\theta_{13}$}

In contrast to the reactor experiments accelerator measurement of 
$\theta_{13}$ is ``contaminated'' (or enriched) by the 
other mixing parameters, in particular by $\delta$ in the case of 
low energy superbeam experiments. 
The sensitivity to $\theta_{13}$ therefore depends upon $\delta$ in a 
significant way. 
Though it sounds like drawback of the accelerator method, it in turn 
means that the LBL $\theta_{13}$ experiments can be upgraded to 
search for leptonic CP violation.
(This is why and how the low-energy superbeam was originally 
motivated in \cite{lowECP}.)
There exist an approved experiment T2K \cite{T2K} using the 0.75 MW 
neutrino beam from J-PARC, and a competitive proposal of 
NO$\nu$A \cite{NOVA} which uses NuMI beam line in Fermilab. 
The sensitivity to $\theta_{13}$, is roughly speaking, up to 
$\sin ^2 2\theta_{13} \sim 0.01$. 
However, the better knowledge of background rejection and the 
systematic errors are required to make the number more solid. 
Though less sensitive, MINOS \cite{MINOS} and OPERA \cite{OPERA} 
have some sensitivities to $\theta_{13}$.

If $\theta_{13}$ is really small, $\sin ^2 2\theta_{13} < 0.01$, 
probably we need new technology to explore the region of $\theta_{13}$. 
The best candidates are neutrino factory \cite{nufact} 
or the beta beam \cite{beta}. 
For them we refer \cite{nufact_eu-us} for overview and for extensive references.

\section{Vacuum vs. matter effects}
\label{vac-matter}

To proceed further, we need some knowledges on neutrino 
oscillation in matter. 
There are several ways to simply understand the matter effect 
in neutrino oscillations. 
One is to use perturbative approach \cite{AKS,MNprd98}. 
The other is to rely on Cervera {\it et al.} formula \cite{golden} 
which applies to higher matter densities. 
The most important reason why we want to understand the feature 
of vacuum-matter interplay in neutrino oscillation is that they tend to 
mix and confuse with each other. 
For the early references which took into account the matter effect which 
inevitably comes in into LBL CP violation search, see e.g., 
\cite{AKS,MNprd98,yasuda,golden}.

\begin{figure}[htbp]
\begin{center}
\vglue 0.3cm
\includegraphics[width=0.6\textwidth]{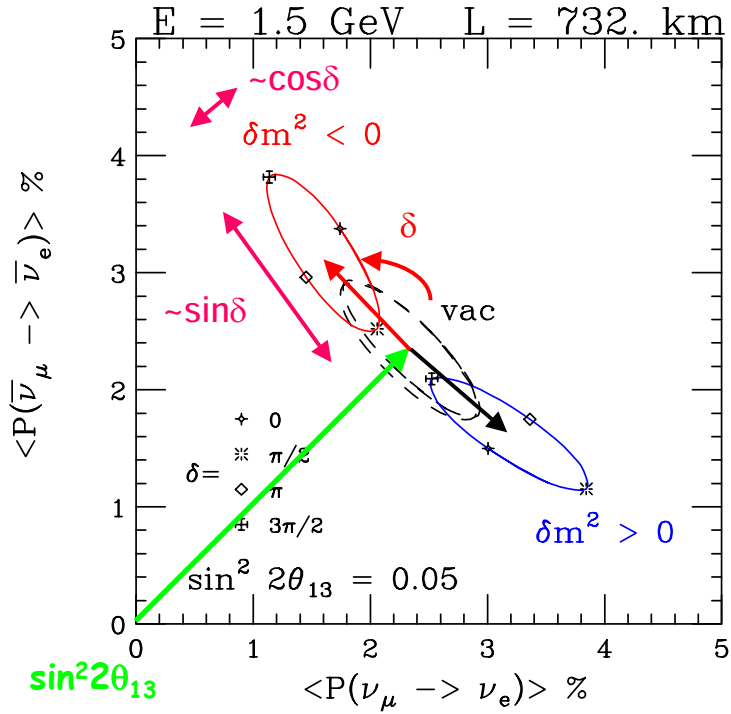}
\end{center}
\caption{
A $P$- $\bar{P}$ bi-probability plot with experimental parameters 
corresponding to NuMI off-axis project is presented for the purpose 
of exhibiting characteristic features of the neutrino oscillations 
relevant for low-energy superbeam experiments. 
Namely, it can disply competing three effects, CP violating and CP 
conserving effects due to $\delta$ as well as the matter effects 
in a compact fashion. For more detailed description of its properties, 
see \protect\cite{MNjhep01}. 
The art work is done by Adam Para. 
}
\label{bi-P}
\end{figure}

Probably, the simplest way to understand the matter effect 
as well as CP phase effect is to rely on the CP trajectory 
diagram in $P(\nu_{\mu} \rightarrow \nu_{e})$ and 
$P(\bar{\nu}_{\mu} \rightarrow \bar{\nu}_{e})$ space, 
for short, the bi-probability plot \cite{MNjhep01}. 
It is given in Fig.~\ref{bi-P}. 
By writing the bi-probability diagram one can easily understand the 
relative importance of CP and the matter effects in a pictorial way; 
Magnitude of effect of CP violating phase $\delta$ is represented 
as the size of the ellipses, while that of the matter effect can be read off 
as a separation between the two ellipses with 
positive (blue ellipse) and negative (red ellipse) sign of $\Delta m^2_{31}$.
The distance from the origin to the ellipse complex represents 
$s^2_{23} \sin^2 2\theta_{13}$. 
We mention that the bi-probability plot can be extended to the one 
which includes T violation in which the charming relations between 
probabilities called the CP-CP and the T-CP relations are hidden 
\cite{MNP1,CPT-unity}.

\section{CP violating phase $\delta$}
\label{delta}

If $\theta_{13}$ is not too small and is within reach by the next 
generation reactor or LBL experiments, the door is open to 
search for leptonic CP violation using conventional superbeam. 
When people started to think about the possibility of observing 
CP violation there were two alternative ways to approach the 
goal, high-energy vs. low-energy options, 
as described in \cite{NOW00_mina}. 
The high-energy option, the majority at that time, is based on the 
idea of neutrino factory \cite{nufact} which utilizes intense neutrino 
beam from muon storage ring. 
Because background can be suppressed to a very small level 
due to clean detection of high-energy muons, the sensitivity to 
$\theta_{13}$ and $\delta$ can be extremely good. 
We do not quote the number here because its re-examination 
by taking into account the possibility of lowering the threshold 
is ongoing in the context of neutrino factory International Scoping 
Study \cite{ISS}, which should be available soon.

The low-energy option is based on very simple fact that the effect 
of CP phase $\delta$ is large at low energies \cite{yasuda,lowECP}. 
What is good in the low-energy option is that it can be realized 
with conventional $\nu_{\mu}$ superbeam. 
It opens the possibility that the CP violation search can be pursued 
by relying on known beam technology with no need of an 
extensive R$\&$D efforts, and is doable in the next 10-15 years 
if we can enjoy generous governmental support. 
On the other hand, $\nu_{e}$ appearance search with 
conventional $\nu_{\mu}$ beam inevitably has the intrinsic problem 
of background, not only of the beam origin but also due to the 
neutral current (NC) $\pi^0$ in the case of water Cherenkov detectors.
Despite the potential difficulties, the possibility of experimental search 
for leptonic CP violation became the realistic option when LOI of the T2K 
experiment with optimistic conclusion was submitted \cite{T2K}.

Unfortunately, the optimism in the early era was challenged by several 
potential obstacles. 
First of all, reducing the systematic error to a required level, 
a few \% level, is a tremendous task. 
Good news is that several experiments are going on, or to be done,  
to measure hadron production \cite{HARP,MIPP} 
and neutrino nucleus interaction cross sections \cite{nuint}. 
Other difficulties include, for example: 
the possibility that CP violation could be masked by the 
unknown sign of $\Delta m^2_{31}$, 
or in more general context the presence of parameter degeneracy 
\cite{intrinsic,MNjhep01,octant} 
which can obscure the CP violation, 
which will be the topics of the next section.

\section{Parameter degeneracy}
\label{Pdege}

Since sometime ago people recognized that measurement 
of $P(\nu_{\mu} \rightarrow \nu_{e})$ and 
$P(\bar{\nu}_{\mu} \rightarrow \bar{\nu}_{e})$ 
at a particular energy, no matter how accurate they are, 
allows multiple solutions of $\theta_{13}$ and $\delta$, 
the problem of parameter degeneracy. 
The nature of the degeneracy can be understood as 
the intrinsic degeneracy \cite{intrinsic}, 
which is duplicated by the unknown sign of $\Delta m^2$ \cite{MNjhep01}, 
and the possible octant ambiguity of $\theta_{23}$ \cite{octant} 
if it is not maximal. 
For an overview of the resultant eight-fold parameter degeneracy, see e.g., 
\cite{BMW,MNP2}.

\begin{figure}[htbp]
\begin{center}
\vspace{0.6cm}
\includegraphics[width=0.6\textwidth]{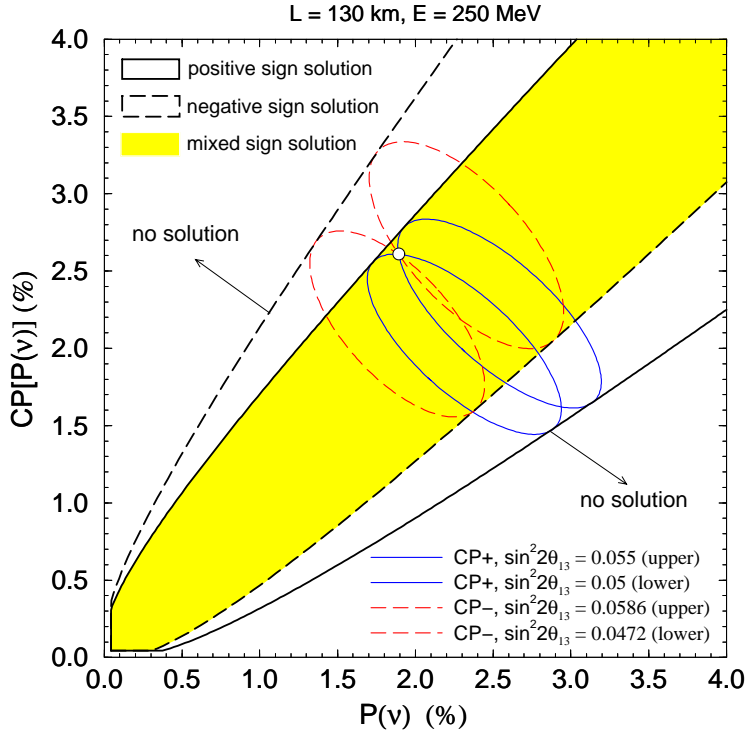}
\end{center}
\caption{
An example of the degenerate solutions for the CERN-Frejius project
in the $P(\nu) \equiv P(\nu_\mu \rightarrow \nu_e)$ verses 
$CP[P(\nu)] \equiv P(\bar{\nu}_\mu \rightarrow \bar{\nu}_e)$
plane.  
Between the solid (dashed) lines is the allowed region for
positive (negative) $\Delta m^2_{31}$ and the shaded region 
is where solution for both signs are allowed. 
The solid (dashed) ellipses are for positive (negative) $\Delta m^2_{31}$
and they all meet at a single point.
This is the CP parameter degeneracy problem.
We have used a fixed neutrino energy of 250 MeV and 
a baseline of 130 km.  
The mixing parameters are fixed to be 
$|\Delta m^2_{13}| = 3 \times 10^{-3} eV^2$,
$\sin^2 2\theta_{23}=1.0$,
$\Delta m^2_{12} = +5 \times 10^{-5} eV^2$,
$\sin^2 2\theta_{12}=0.8$ and
$ Y_e \rho  = 1.5$ g cm$^{-3}$. 
The figure is taken from \cite{MNP2}. 
}
\label{degeneracy}
\end{figure}

It is in fact easy to understand the cause of the parameter degeneracy 
if we use the bi-probability plot. 
Look at Fig.~\ref{degeneracy}. 
Suppose that your experimentalist friend gives you the measurement 
point denoted as an open circle in Fig.~\ref{degeneracy}. 
Then, it is evident that you can draw two ellipses, 
as shown in blue solid lines in Fig.~\ref{degeneracy}, 
that pass through the observed point, which implies the existence of 
two solutions of $\theta_{13}$ and $\delta$. 
The two-fold ambiguity is usually called the intrinsic degeneracy. 
If we are ignorant of the neutrino mass hierarchy, i.e., 
the sign of $\Delta m^2_{31}$, the two more ellipses can be drawn, 
as shown by the red dashed line in Fig.~\ref{degeneracy}; 
duplication of the solution by the unknown mass hierarchy. 
Altogether one has four-fold parameter degeneracy.

Unfortunately, it is not the end of the story.
If $\theta_{23}$ is not maximal, we are enriched with another degeneracy, 
the $\theta_{23}$  octant degeneracy.
The $\nu_{\mu}$ disappearance measurement of 
$P(\nu_{\mu} \rightarrow \nu_{\mu})$ gives a value of $\sin^2 2\theta_{23}$. 
It then allows two solutions of $\theta_{23}$ if $\theta_{23} \neq \pi/4$, 
$s^2_{23} = \frac{1}{2} \left[ 1 \pm \sqrt{1- \sin^2{2\theta_{23}}} \right]$, 
one in the first octant and the 
other in the second octant. 
Since this is ``orthogonal'' to the intrinsic and the sign $\Delta m^2_{31}$ 
degeneracies with four solutions of $\theta_{13}$ and $\delta$, 
the total eight-fold degeneracy results.

Prior to the systematic discussion of how to solve the degeneracy 
we want to mention about the simplest method of solving 
the $\theta_{13} - \delta$ degeneracy. 
By tuning the beam energy to the ``shrunk ellipse limit'' \cite{KMN02} 
the degeneracy can be reduced to the two-fold one, 
$\delta \leftrightarrow \pi - \delta$. 
Notice that there is no confusion between CP violation 
and CP conservation even in the presence of this degeneracy.

\section{Resolving the eight-fold parameter degeneracy}
\label{resolveD}

It is known that degeneracy of neutrino oscillation parameters 
acts as a severe limiting factor to the precision determination 
of $\theta_{13}$, $\theta_{23}$, and $\delta$. 
It is particularly true for the $\theta_{23}$ octant degeneracy 
\cite{MSS04}. 
Expecting the era of precision measurement in the next 10-30 years, 
it is the time that the formulation of the well defined strategy for 
exploring the whole structure of the lepton flavor mixing is of immense need.

Toward the goal, I explain in detail how the degeneracy can be resolved 
by using a concrete setting, which is called ``T2KK''. 
It is an acronym for Tokai-to-Kamioka-Korea two detector complex,
an upgraded next project to T2K phase I for exploring 
the whole structure of lepton flavor mixing \cite{T2KK1st,T2KK2nd}.
It utilizes two half a megaton (0.27 Mton fiducial volume) water 
Cherenkov detectors one in Kamioka (295 km) and the other in 
somewhere in Korea ($\sim$1000 km) which receive 
$\nu_{\mu}$ and $\bar{\nu}_{\mu}$ superbeam of 4 MW from 
J-PARC facility. We assume 4 years of running with each neutrino and 
antineutrino mode.  
For further details of T2KK, please consult to the original manuscripts 
\cite{T2KK1st,T2KK2nd}. 
For a broader view of T2KK including wider class of detector options 
and locations, see the web page of the workshops 
which are focused on this project \cite{T2KKweb}. 
Though T2KK is {\it not} the unique way of resolving the eight-fold 
parameter degeneracy it is nice to have a concrete project 
to solve all the degeneracy {\it in situ}; 
It provides the bottom line understanding on how it can be lifted, 
and the lesson may be useful to think about alternative ways.

How does T2KK solve the 8-fold parameter degeneracy? 
In a nutshell, the setting can resolve the three kind of degeneracies 
in the following ways:

\begin{itemize}

\item

The intrinsic degeneracy; 
Spectrum information solves the intrinsic degeneracy.

\item

The sign-$\Delta m^2$ degeneracy; 
Difference in the earth matter effect between the intermediate (Kamioka) 
and the far (Korea) detectors solves the sign-$\Delta m^2$ degeneracy. 

\item

The $\theta_{23}$ octant degeneracy; 
Difference in solar $\Delta m^2$ oscillation effect 
(which is proportional to $c^2_{23}$) between the intermediate 
and the far detectors solves the $\theta_{23}$ octant degeneracy. 

\end{itemize}

\noindent
Let me explain these points one by one.

\subsection{Intrinsic degeneracy}
\label{intrinsic}

\begin{figure}[htbp]
\begin{center}
\vglue 0.3cm
\includegraphics[width=0.46\textwidth]{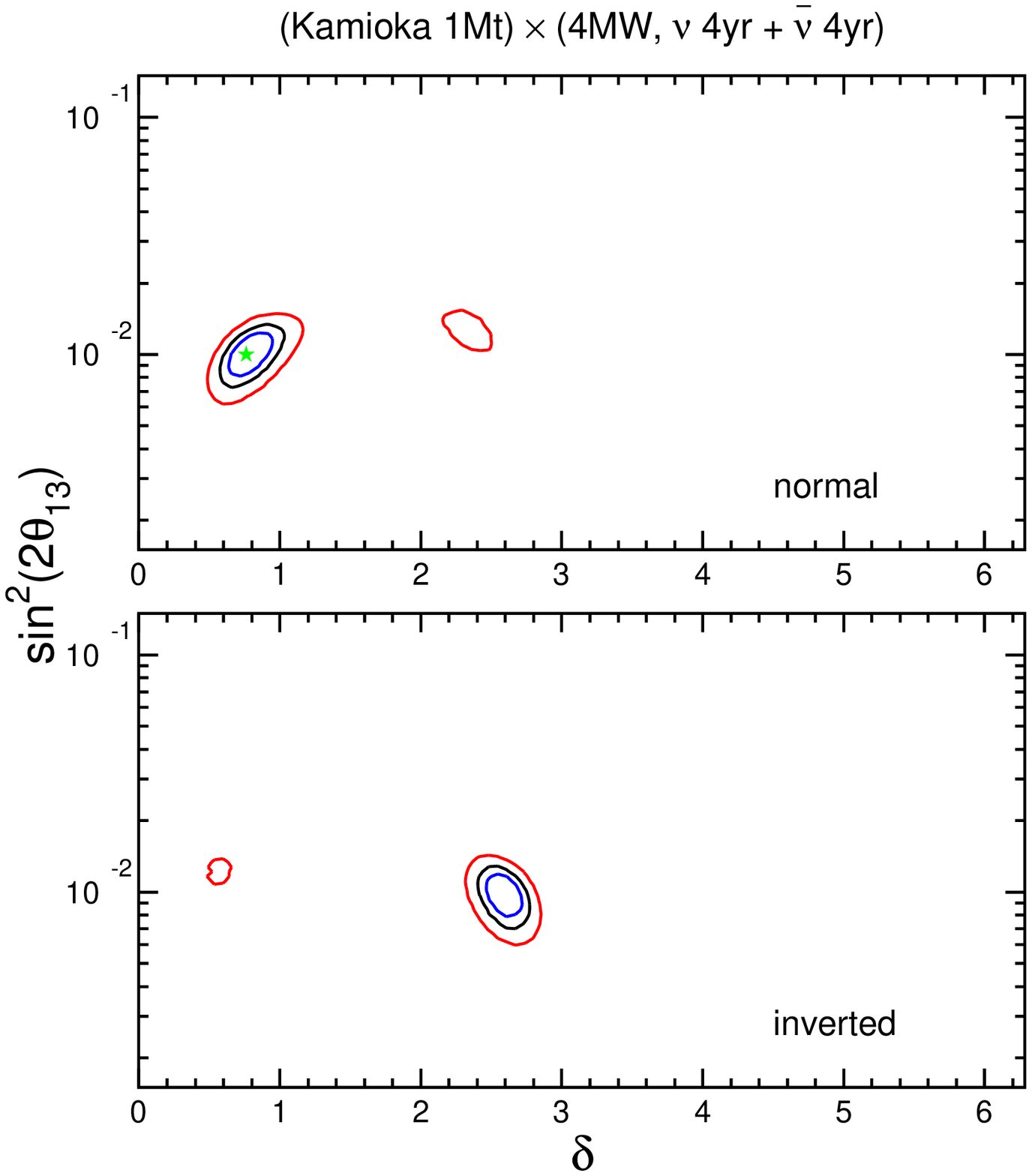}
\includegraphics[width=0.46\textwidth]{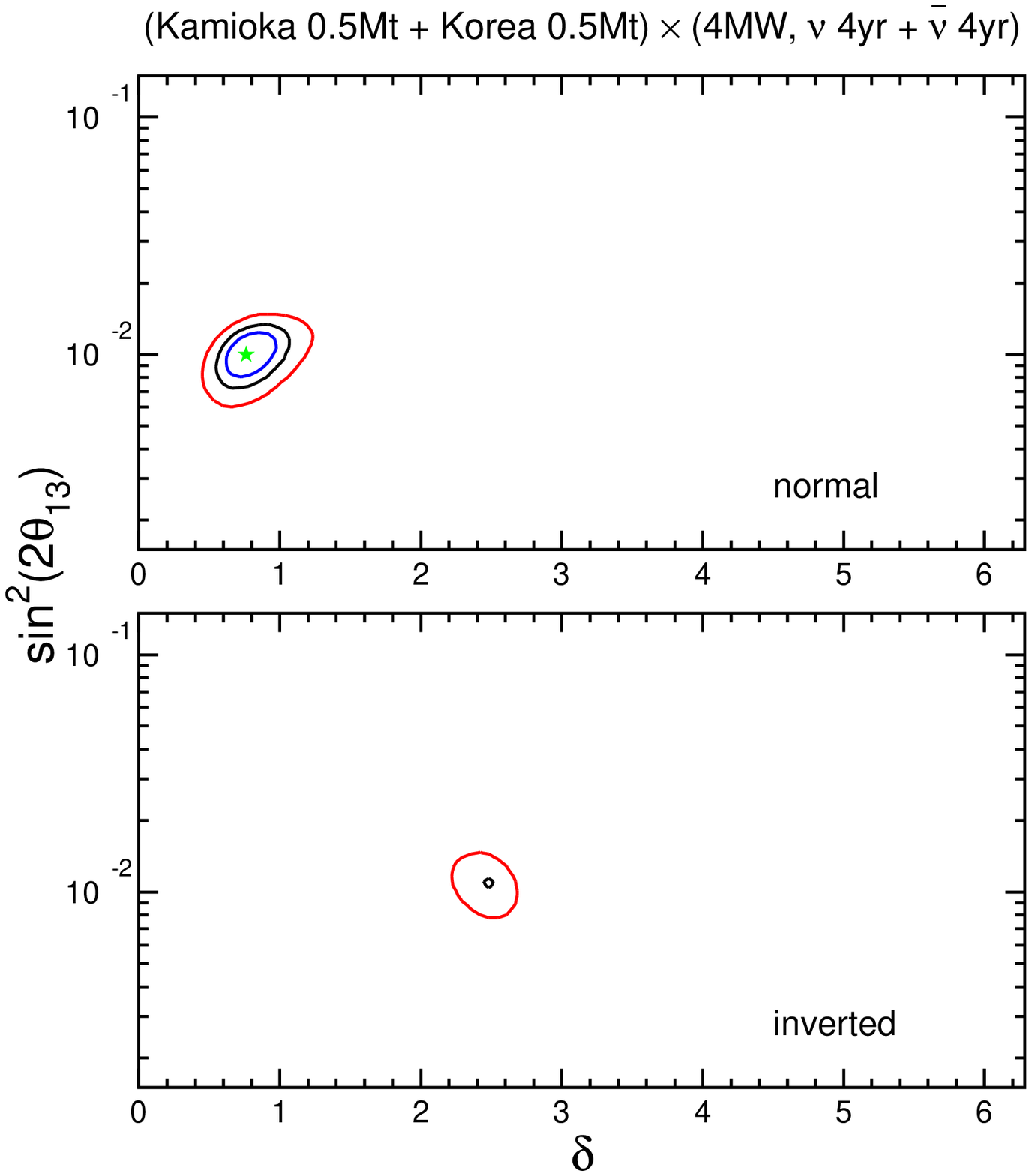}
\end{center}
\caption{The region allowed in $\delta-\sin^2 2\theta_{13}$ space by 4 years of 
neutrino and antineutrino running in  T2K II (left panel), and the 
Kamioka-Korea two detector setting (right panel). 
They are taken from the supplementary figures behind the 
reference \protect\cite{T2KK1st} to which the readers are referred 
for details of the analysis. 
Notice that the standard setting in T2K II, 2 (6) years of 
neutrino (antineutrino) running, leads to a very similar results 
(as given in \cite{NOVE06_mina}) to the one presented in the left 
panel of this figure.  
The true solutions are assumed to be located at 
($\sin^2{2\theta_{13}}$ and $\delta$) = (0.01, $\pi/4$) 
with positive sign of $\Delta m_{31}^2$, as indicated as the green star. 
The intrinsic and the $\Delta m_{31}^2$-sign clones appear in the 
same and the opposite sign $\Delta m_{31}^2$ panels, respectively. 
Three contours in each figure correspond to
the 68\% (blue line), 90\% (black line) and 99\% 
(red line) C.L. sensitivities, respectively.
}
\label{intrinsicKamKorea}
\end{figure}

First look at Fig.~\ref{intrinsicKamKorea} in which 
the sensitivities for resolving the intrinsic degeneracy by 
the Tokai-to-Kamioka phase II (T2K II) setting \cite{T2K} (left panel)
and the Kamioka-Korea two detector setting  (right panel)
are presented. 
Figure \ref{intrinsicKamKorea} is taken from supplementary figures 
prepared for the study reported in \cite{T2KK1st}, in which the details 
of the analysis are described. 
In the left panel of Fig.~\ref{intrinsicKamKorea} it is shown that the 
intrinsic degeneracy in (assumed) each mass hierarchy is almost 
resolved by the T2K II setting at the particular set of values of the 
mixing parameters as indicated in the caption. 
Since the matter effect plays minor role in the T2K II setting it is likely 
that the spectral information is mainly responsible for lifting the intrinsic 
degeneracy.

In the right panel of the same figure it is exhibited that the intrinsic 
degeneracy is completely resolved by the T2KK setting 
at the same values of the mixing parameters, 
indicating power of the two detector method \cite{MN_2detector}. 
Namely, the comparison between the intermediate and the far 
detectors placed at the first and the second oscillation maxima, 
respectively, supersedes a single detector measurement in 
Kamioka with the same total volume in spite of much less 
statistics in the Korean detector.

\subsection{Sign-$\Delta m^2$ degeneracy}
\label{signDm2}

\begin{figure}[htbp]
\begin{center}
\vglue 0.3cm
\includegraphics[width=0.46\textwidth]{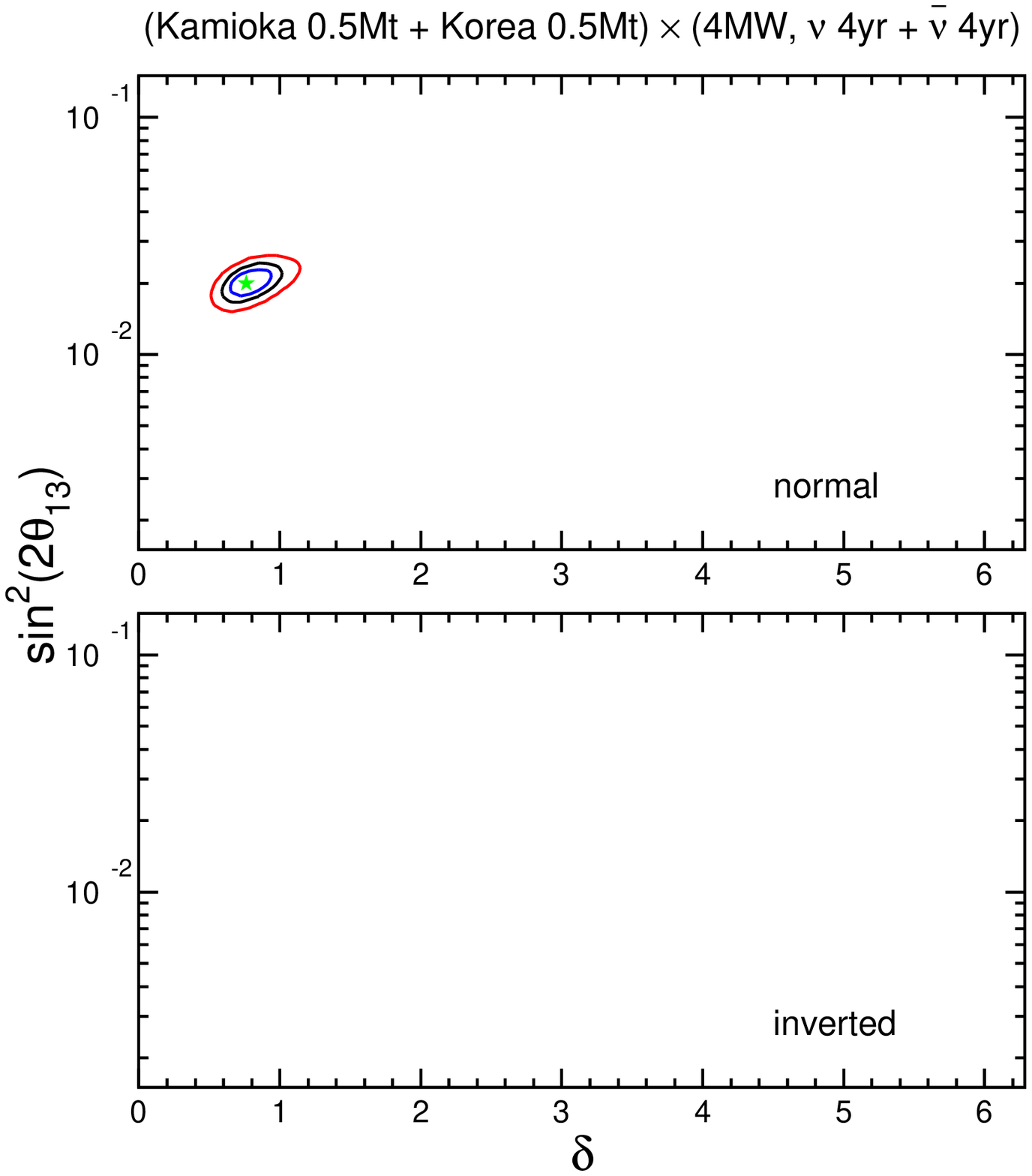}
\includegraphics[width=0.46\textwidth]{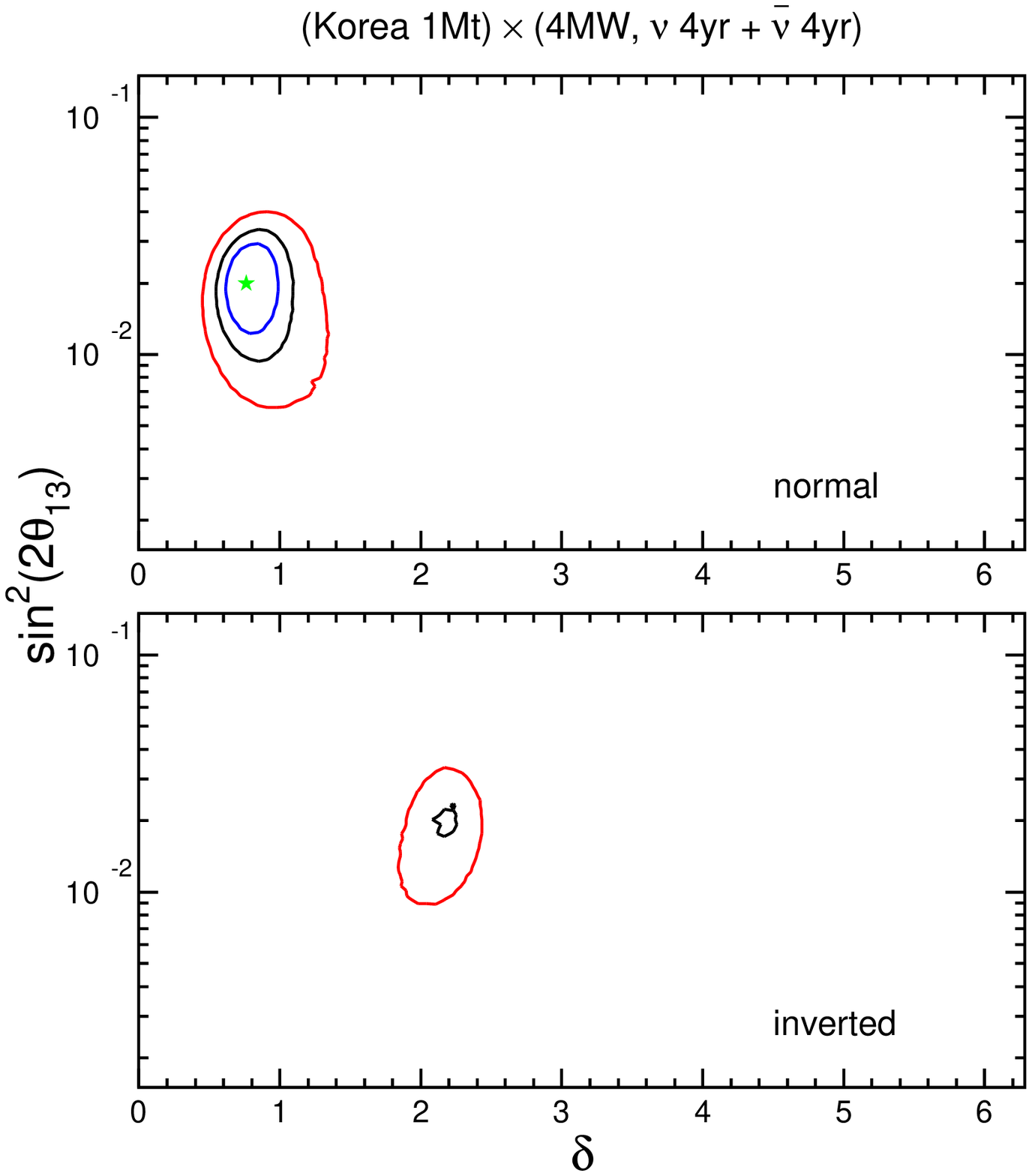}
\end{center}
\caption{The similar sensitivity plot as in Fig.~\ref{intrinsicKamKorea}. 
Left panel is for T2KK and the right panel for a single 0.54 megaton 
detector placed in Korea.
}
\label{signdm2KamKorea}
\end{figure}

It should be noted that the sign-$\Delta m^2$ degeneracy is also lifted 
though incompletely by the Kamioka-Korea setting 
as indicated in the right-lower panel of Fig.~\ref{intrinsicKamKorea}. 
It is well known that the interference effect between the vacuum 
and the matter effects depends upon the mass hierarchy, i.e., 
the sign of $\Delta m^2_{31}$, and many people have been proposed 
to utilize it to resolve the mass hierarchy.

But, it is not the whole story here. 
To indicate this point the sensitivities for the two settings are compared 
in Fig.~\ref{signdm2KamKorea}. 
One is T2KK (left panel) and the other is the case of single 0.54 
megaton detector placed in Korea (right panel). 
It should be noticed that a single detector in Korea with the same 
total volume fails to resolve the sign-$\Delta m^2$ degeneracy 
which is completely lifted by T2KK at the particular values of the 
mixing parameters. 
Again, the sensitivity is enhanced by comparing the yields of the 
two identical detectors.

\begin{figure}[htbp]
\begin{center}
\includegraphics[width=0.90\textwidth]{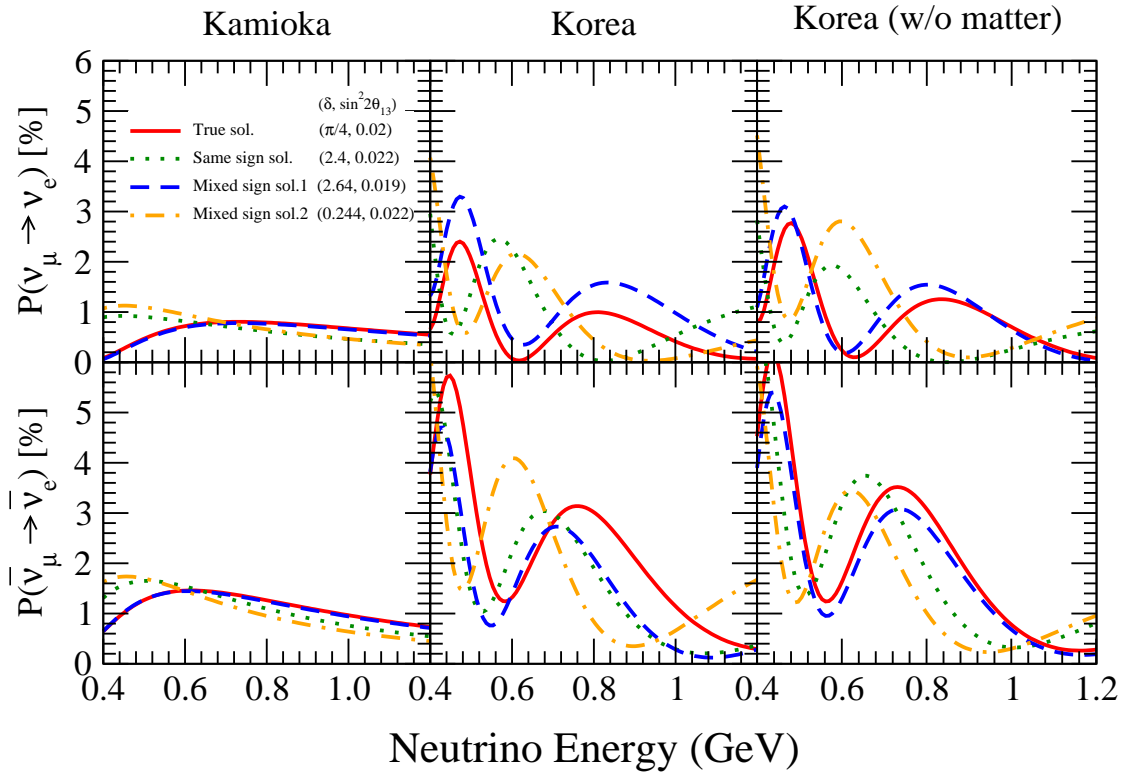}
\end{center}
\vglue -0.3cm
\caption{Neutrino oscillation probabilities corresponding to a 
four-fold degenerate solutions obtained by measurement in 
Kamioka by the rate only analysis are plotted as a function of 
neutrino energy. 
Left panels: appearance probabilities in Kamioka. 
Middle panels: appearance probabilities in Korea.
Right panels: appearance probabilities in Korea, but without the matter
effect.
}
\label{spectral}
\end{figure}

The fact that the T2KK setting can resolve the four-fold degeneracy 
by the spectrum analysis and comparison between the two detectors 
is explained by plotting the energy dependences of the appearance 
probabilities in Fig.~\ref{spectral}.

\subsection{$\theta_{23}$ octant degeneracy}
\label{octant}

The $\theta_{23}$ octant degeneracy arise because accelerator 
disappearance and appearance measurement determine 
$\sin^2{2 \theta_{23}}$ and the combination 
$s^2_{23} \sin^2{2 \theta_{13}}$, respectively, but not 
$s^2_{23}$ itself. Therefore, it is hard to resolve in the 
accelerator experiments using conventional $\nu_{\mu}$ beam. 
See \cite{resolve23} for an explicit analytic treatment of this point.

\begin{figure}[htbp]
\begin{center}
\includegraphics[width=0.7\textwidth]{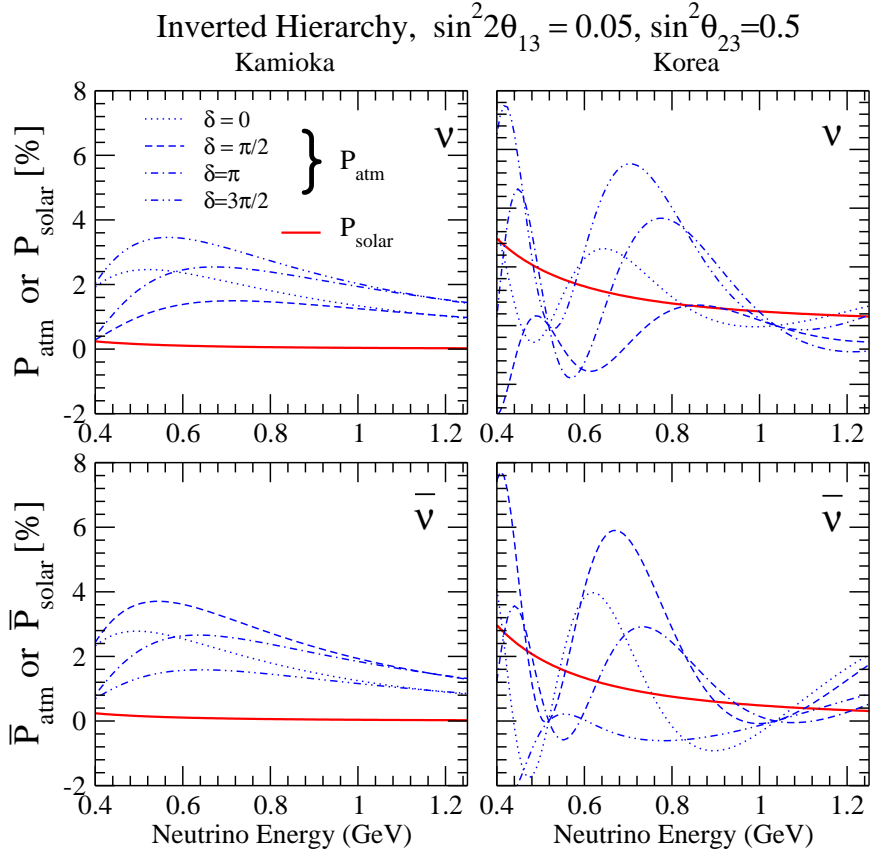}
\end{center}
\caption{
The energy dependence of the solar term (red solid line) is 
contrasted with the ones of atmospheric plus interference 
terms in the $\nu_{e}$ 
appearance oscillation probabilities with various values of CP phase 
$\delta$; 
$\delta=0$ (dotted line), 
$\delta=\pi/2$ (dashed line), 
$\delta=\pi$ (dash-dotted line), and 
$\delta=3\pi/2$ (double-dash-dotted line). 
The neutrino mass hierarchy is assumed to be the inverted one. 
For the corresponding figure of the normal mass hierarchy, see 
\protect\cite{T2KK2nd}.
}
\label{solar-vs-atm}
\end{figure}

One way to solve the $\theta_{23}$ octant degeneracy is to utilize 
the solar $\Delta m^2$ oscillation term. This is the principle used 
by the atmospheric neutrino method for resolving the octant degeneracy 
\cite{atm23,concha-smi_23,choubey}. 
Since the solar term, the last term in Eq.~(\ref{Pmue}), depend upon 
$c^2_{23}$ (not $s^2_{23}$) the degeneracy can be lifted. 
The next question is if it can be distinguished from the rest of 
the atmospheric oscillation terms. Fortunately, the answer seems 
to be yes because of the clear difference in energy dependence, 
as shown in Fig.~\ref{solar-vs-atm}. 
Note that the figure is the inverted hierarchy version of 
Fig.~2 in \cite{T2KK2nd}, and behavior of the solar term compared 
to the atmospheric ones is very similar to in the case of the normal hierarchy.

\subsection{Decoupling between the degeneracies}
\label{decoupling}

In passing, we briefly comment on the problem of decoupling between the 
degeneracies. For a fuller treatment, see \cite{T2KK2nd}. 
The question is as follows: 
People sometimes discuss how to solve the degeneracy A without worrying 
about the degeneracy B, and vise versa. Is this a legitimate procedure? 
We want to answer to this question in the positive under the environment 
that the matter effect can be treated as a perturbation.

To resolve the degeneracy one has to distinguish 
between the values of the oscillation probabilities with the two different 
solutions corresponding to the degeneracy. 
We define the probability difference 
\begin{eqnarray}
\Delta P^{ab}(\nu_{\alpha} \rightarrow \nu_{\beta}) 
&\equiv&  
P \left( \nu_{\alpha} \rightarrow \nu_{\beta}; \theta_{23}^{(a)}, 
\theta_{13}^{(a)}, \delta^{(a)}, (\Delta m^2_{31})^{(a)} \right) 
\nonumber \\
&-& 
P \left( \nu_{\alpha} \rightarrow \nu_{\beta}; \theta_{23}^{(b)}, 
\theta_{13}^{(b)}, \delta^{(b)}, (\Delta m^2_{31})^{(b)} \right), 
\label{DeltaPdef}
\end{eqnarray}
as a measure for it 
where the superscripts $a$ and $b$ label the degenerate solutions. 
Suppose that we are discussing the degeneracy A. 
The decoupling between the degeneracies A and B  
holds if $\Delta P^{ab}$ defined in (\ref{DeltaPdef}) for the degeneracy A 
is invariant under the replacement of the mixing parameters 
corresponding to the degeneracy B, and vice versa.


The best example of the decoupling is given by the one between 
the $\theta_{23}$ octant and the sign-$\Delta m^2$ degeneracies. 
Therefore, let us describe it here, leaving discussions on other cases 
to \cite{T2KK2nd}. One can easily compute 
$\Delta P^{1st-2nd}(\nu_{\mu} \rightarrow \nu_{e})$ for the 
$\theta_{23}$ octant degeneracy by using (\ref{Pmue}). 
It consists of the solar and the solar-atmospheric interference terms, 
with over-all factor of $\cos{2\theta_{23}}$ because of the property 
$J_r^{1st} - J_r^{2nd} = \cos{2\theta_{23}^{1st}} J_r^{1st}$ 
in leading order in $\cos{2\theta_{23}}$.
The remarkable feature of 
$\Delta P^{1st-2nd}(\nu_{\mu} \rightarrow \nu_{e})$ 
is that the leading-order matter effect terms drops out completely.

Now, we notice the key feature of $\Delta P^{1st-2nd}(\nu_{\mu} \rightarrow \nu_{e})$;   
It is invariant under the transformations 
$\Delta m^2_{31}) \rightarrow - \Delta m^2_{31}$ and 
$\delta \rightarrow \pi - \delta$, 
which exchanges the two sign-$\Delta m^2$ degenerate solutions, 
the invariance which holds in the presence of the solar term. 
It means that resolution of the $\theta_{23}$ degeneracy can be 
executed without knowing the mass hierarchy in experimental 
set up which allows perturbative treatment of matter effect.

Next, we examine the inverse problem; 
Does the determination of mass hierarchy decouple from resolution of 
the $\theta_{23}$ degeneracy?
One can compute in the similar way $\Delta P^{norm-inv}$ for 
the sign-$\Delta m^2$ degeneracy. 
Because the exchange of two sign-$\Delta m^2$ degenerate solutions 
is the approximate symmetry of the vacuum oscillation probability \cite{MNjhep01}, 
most of the vacuum terms drops out.
We observe that $\Delta P^{norm-inv}(\nu_{\mu} \rightarrow \nu_{e})$ 
is invariant under transformation 
$\theta_{23}^{1st} \rightarrow \theta_{23}^{2nd}$ and 
$\theta_{13}^{1st} \rightarrow \theta_{13}^{2nd}$, 
because its $\theta_{13}$ and $\theta_{23}$ dependences 
are through the combination 
$\sin^2{2\theta_{13}} s_{23}^{2}$. 
Therefore, resolution of the mass hierarchy can be carried out 
independently of which solution of the $\theta_{23}$ degeneracy 
is realized in nature.

We mention here that the decoupling argument can be generalized 
to include the other pair of degeneracies as done in \cite{T2KK2nd}.

\subsection{Analysis results}

\begin{figure}[htbp]
\vglue 0.3cm
\begin{center}
\includegraphics[width=0.64\textwidth]{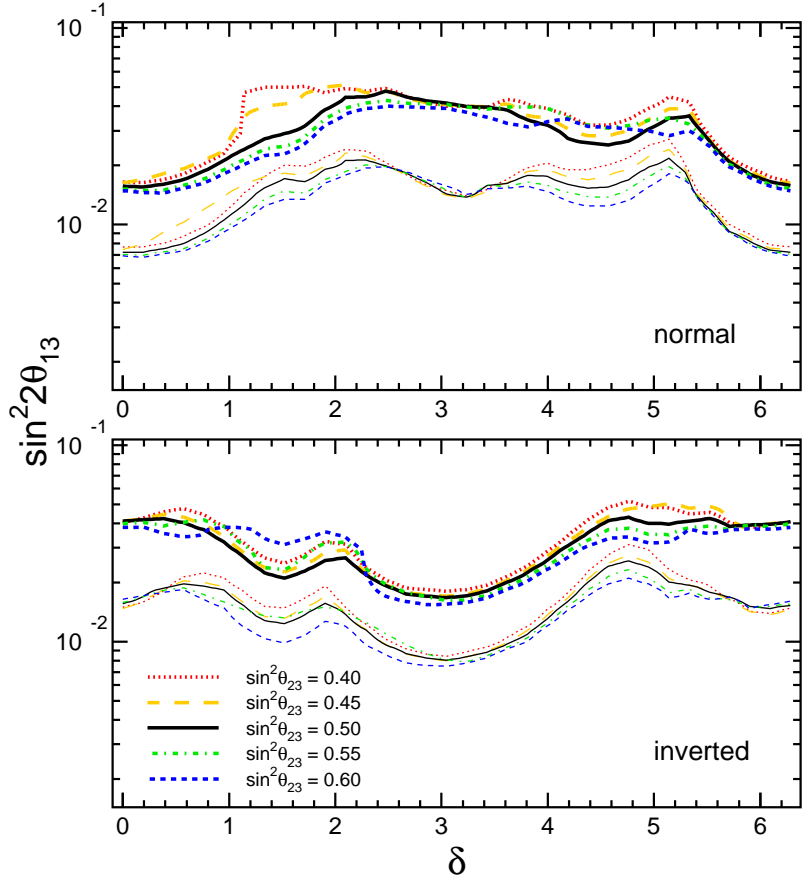}
\end{center}
\caption{ 2(thin lines) and 3(thick lines) 
standard deviation sensitivities to the mass hierarchy 
determination for
several values of $\sin ^2 2 \theta_{23}$ 
(red dotted, yellow long-dashed, black solid, green dash-dotted, 
and blue short-dashed lines show the results for 
$\sin^2 \theta_{23} =$ 0.40, 0.45, 0.50, 0.55 and 0.60, 
respectively). The sensitivity is defined in the plane
of $\sin ^2 2 \theta_{13}$ versus CP phase $\delta$.
The top and bottom panels show the cases for positive and 
negative mass hierarchies, respectively. 
Taken from \cite{T2KK2nd}. 
}
\label{sensitivity-mass-hierarchy}
\end{figure}

\begin{figure}[htbp]
\vglue 0.3cm
\begin{center}
\includegraphics[width=0.64\textwidth]{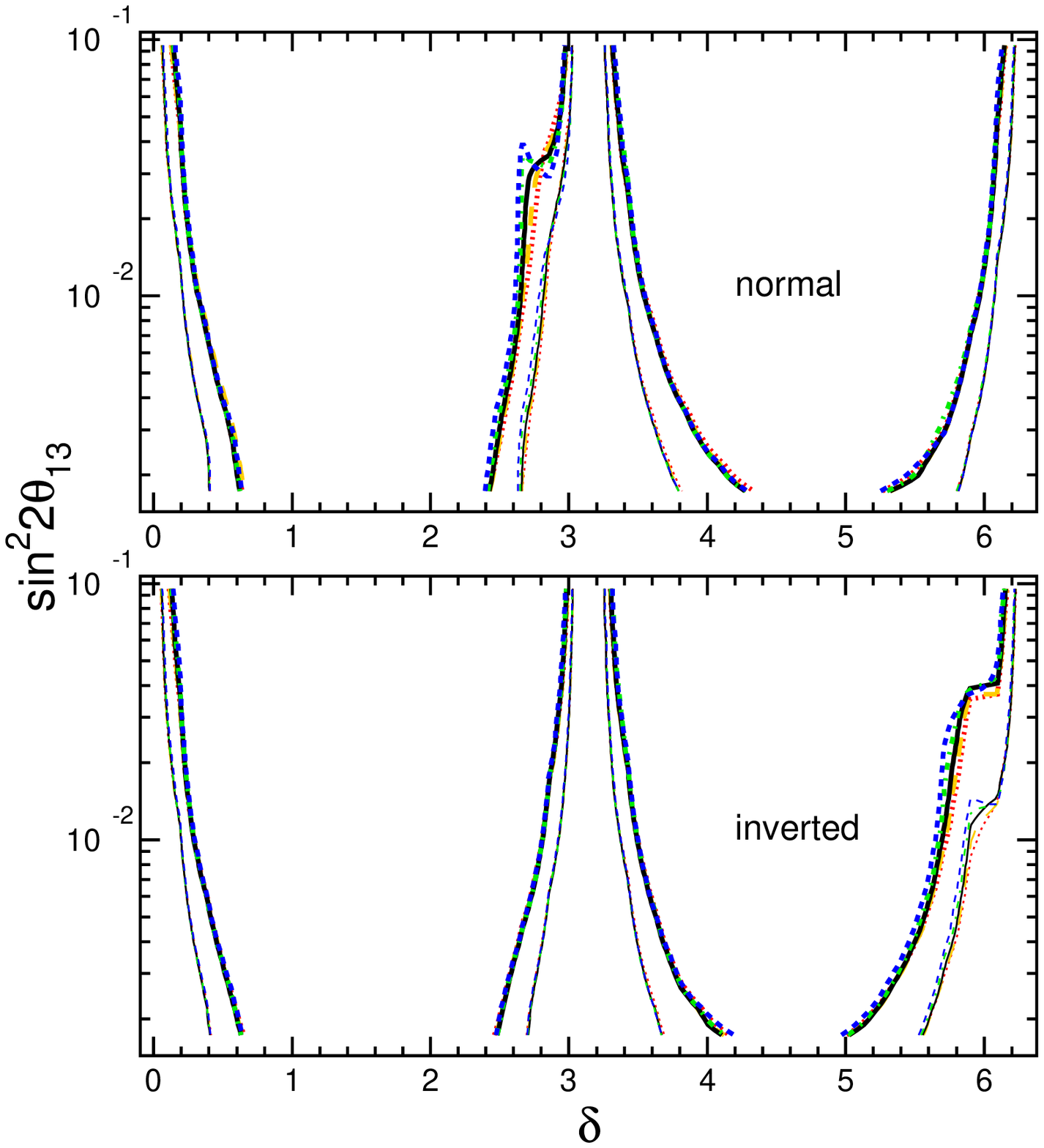}
\end{center}
\vglue -0.2cm
\caption{ Sensitivities to the CP violation, 
$\sin \delta \neq 0$. 
The meaning of the lines 
and colors are identical to that in 
Fig.~\ref{sensitivity-mass-hierarchy}.
Taken from \cite{T2KK2nd}. 
}
\label{sensitivity-CP}
\end{figure}

Since the space is quite limited, we directly go to the results of our 
analysis. 
The original analysis in \cite{T2KK1st} has been 
re-examined with an improved code which takes into account a 
difference between beam profiles in the intermediate and the far 
detectors, and the inclusion of the muon disappearance channel \cite{T2KK2nd}. 
In Fig.~\ref{sensitivity-mass-hierarchy}, and in Fig.~\ref{sensitivity-CP}, 
the results of re-analysis for the mass hierarchy resolution 
and CP violation, respectively, are presented. 
Figure \ref{sensitivity-mass-hierarchy} shows that the sensitivity to 
the mass hierarchy depends very weakly to $\theta_{23}$, 
as expected by the decoupling argument given in \cite{T2KK2nd}. 
The same argument suggests that they obey the scaling behavior; 
the curves falls to a single curve if plotted by $s^2_{23} \sin ^2 2\theta_{13}$. 
The sensitivity greatly improves the one possessed by the original 
T2K II setup and is competitive to other similar projects. 
See \cite{T2KK1st} for comparison between the performances of 
T2KK and T2K II setting.

The $\theta_{23}$-independence of the CP sensitivity is even more 
prominent, as shown in Fig.~\ref{sensitivity-CP}. 
This feature is again consistent with the decoupling argument. 
The sensitivity to CP violation is similar to that of the T2K II setting 
except for at large $\theta_{13}$ region where the T2KK sensitivity 
surpasses that of the T2K II. 
It is due to the fact that the identical two-detector setting solves the 
degeneracies. 
We emphasize that the CP sensitivity of T2KK setting at the large 
$\theta_{13}$ region seems to be the largest among the similar 
proposals including neutrino factory.

\begin{figure}[htbp]
\vglue 0.3cm
\begin{center}
\includegraphics[width=0.78\textwidth]{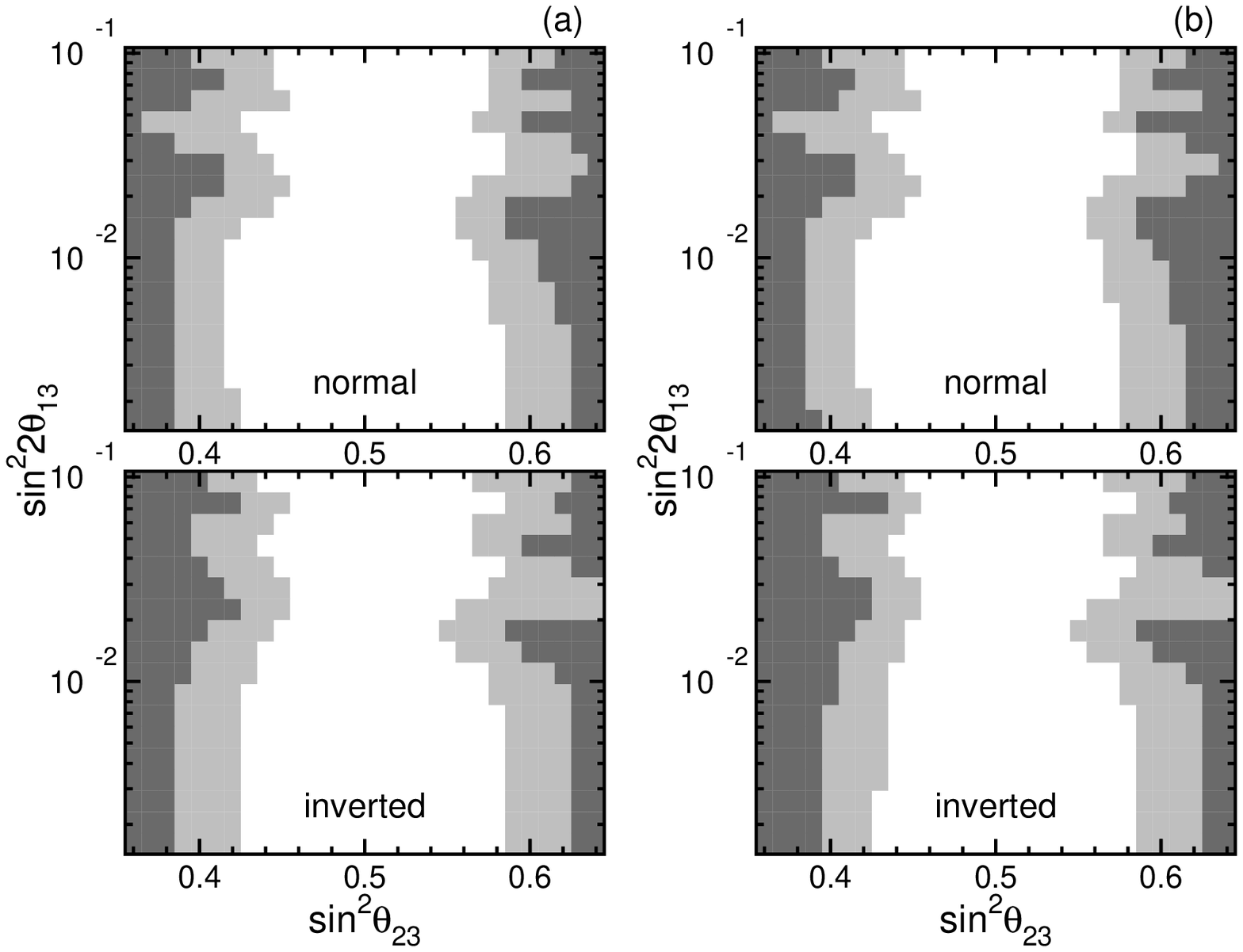}
\end{center}
\vglue -0.2cm
\caption{ 2 (light gray area) and 3 (dark gray area) 
standard deviation sensitivities to the $\theta_{23}$
octant degeneracy for 0.27~Mton detectors both in Kamioka
and Korea.
4 years running with neutrino beam and another 4 years with 
anti-neutrino beam are assumed.
In (a), the sensitivity is defined so that the experiment 
is able to identify the octant of $\theta_{23}$ for any 
values of the CP phase $\delta$. In (b), it is defined so that the 
experiment is able to identify the octant of $\theta_{23}$ 
for half of the CP $\delta$ phase space.
Taken from \cite{T2KK2nd}. 
}
\label{sensitivity-theta23-octant}
\end{figure}

In Fig.~\ref{sensitivity-theta23-octant}, the sensitivity to the 
$\theta_{23}$ octant degeneracy is presented. 
From this figure, we conclude that the experiment 
we consider here is able to solve the octant ambiguity,
if $\sin^2 \theta_{23} < 0.38\,(0.42) $ or $>0.62\,(0.58)$ at 3
(2) standard deviation CL. 
Roughly speaking, the sensitivity is independent of $\theta_{13}$ 
and the mass hierarchy. 
The dependence of this sensitivity on the CP phase $\delta$ is a 
mild one as one can see by comparing the left and the right panels 
of Fig.~\ref{sensitivity-theta23-octant}, providing another evidence 
for decoupling.

As discussed in detail in \cite{resolve23}, the $\theta_{23}$ degeneracy 
is the difficult one to solve only by the accelerator experiments. 
Though the argument is still true, T2KK circumvents it because it has 
sensitivity to the solar term. 
Yet, the sensitivity is quite limited if plotted in $s^2_{23}$ plane, 
as one can observe in Fig.~\ref{sensitivity-theta23-octant}. 
Nonetheless, we stress that it is not easy to supersede the 
sensitivity presented in Fig.~\ref{sensitivity-theta23-octant}. 
For example, T2KK's sensitivity is slightly better than the one by 
the atmospheric neutrino method based on 3 years observation in 
Hyper-Kamiokande reported in \cite{atm23}.

We emphasize that our estimates of sensitivities for the mass hierarchy 
resolution, CP violation, and the $\theta_{23}$ octant degeneracy 
are based on the known technology for 
rejecting NC induced background in water Cherenkov detectors. 
Moreover, we have used a conservative value of 5\% for most of 
the systematic errors \cite{T2KK1st,T2KK2nd}. 
Therefore, our results can be regarded as robust bottom-line 
sensitivities achievable by conventional superbeam experiments. 
Of course, there may be ways to improve the sensitivities over the 
current T2KK design.

At the end of this section, we should mention that the method 
explored in this article is by no means unique. 
With regard to the sign-$\Delta m^2$ (mass hierarchy) degeneracy 
we note that other methods include the one which utilize 
atmospheric neutrinos \cite{atm-method}, 
supernova neutrinos \cite{SN-method1,SN-method2}, 
neutrino-less double beta decay \cite{double-beta}, and 
$\nu_{e}$ and $\nu_{\mu}$ disappearance channels \cite{GJK,NPZ,MNPZ}. 
We have already mentioned about the $\theta_{23}$ octant degeneracy, 
and a further comment follows immediately below.

\section{Reactor-Accelerator method for $\theta_{23}$ octant degeneracy}
\label{reactor-accelerator}

\begin{figure}[htbp]
\begin{center}
\hglue 0.8cm
\includegraphics[width=0.86\textwidth]{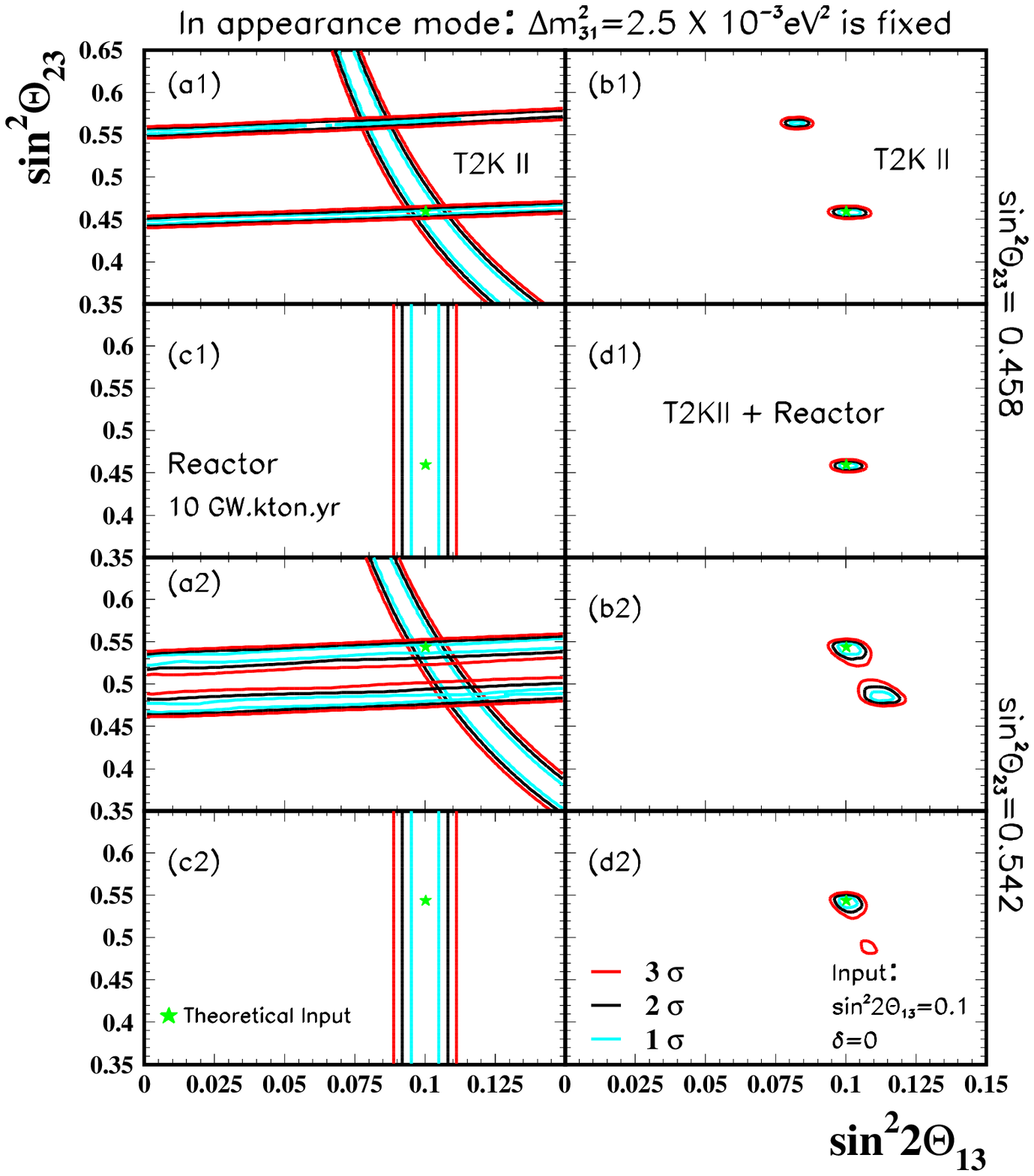}
\end{center}
\vglue -1.5cm
\caption{
The upper (lower) four panels describe the process of how the $\theta_{23}$ 
octant degeneracy can be resolved for the case where the true value 
of $\sin^2\theta_{23}$ = 0.458 (0.542), corresponding to 
$\sin^2{2\theta_{23}}=0.993$. The other input mixing
parameters are given as 
$\Delta m^2_{31} = 2.5\times 10^{-3}$ eV$^2$, 
$\sin^22 \theta_{13}$ = 0.1 and $\delta =0$, 
$\Delta m^2_{21} = 8.0\times 10^{-5}$ eV$^2$, 
$\sin^2 \theta_{12}$ = 0.31 (the input values of 
$\sin^22 \theta_{13}$ and $\sin^2\theta_{23}$ 
are indicated by the symbol of star in the plot).
(a) The regions enclosed by the solid and the dashed curves are 
allowed regions only by the results of appearance and disappearance 
accelerator measurement, respectively. 
(b) The regions that remain allowed when results of appearance 
and disappearance measurement are combined. 
(c) The regions  allowed by reactor measurement.
(d) The regions allowed after combining the results of appearance and
disappearance accelerator experiments with the reactor measurement.
The exposures for accelerator are assumed to be 2 (6) years of neutrino 
(anti-neutrino) 
running with 4 MW beam power with Hyper-Kamionande whose fiducial 
volume is 0.54 Mt, whereas for the reactor we assume an exposure of 
10 GW$\cdot$kt$\cdot$yr. 
The case of optimistic systematic error is taken. 
The figure is taken from \cite{resolve23}. 
}
\label{23fig1}
\end{figure}

\begin{figure}[htbp]
\vglue -2.0cm
\begin{center}
\hglue 2cm
\includegraphics[width=0.92\textwidth]{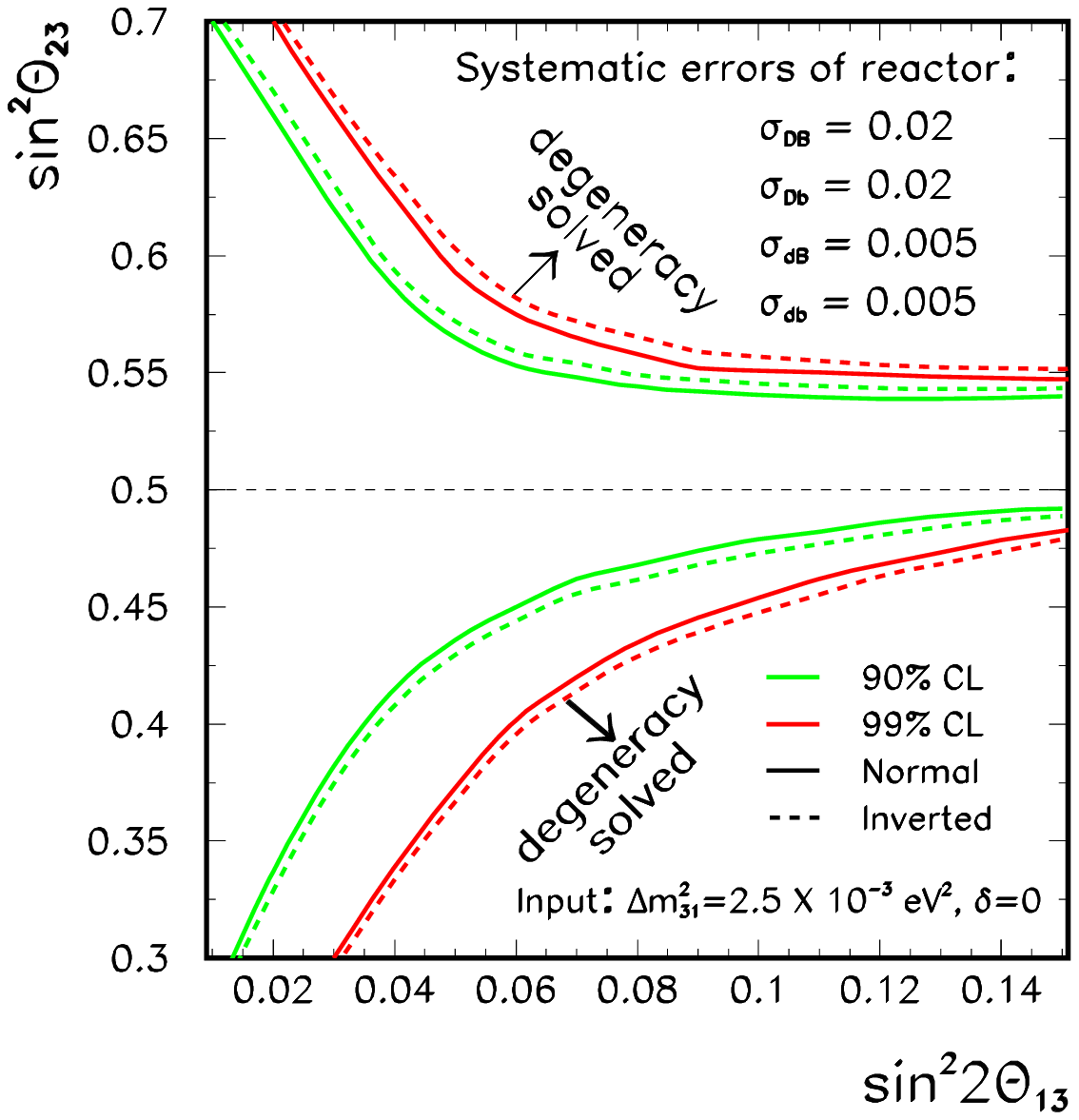}
\vglue -5.4cm
\hglue 2cm
\includegraphics[width=0.92\textwidth]{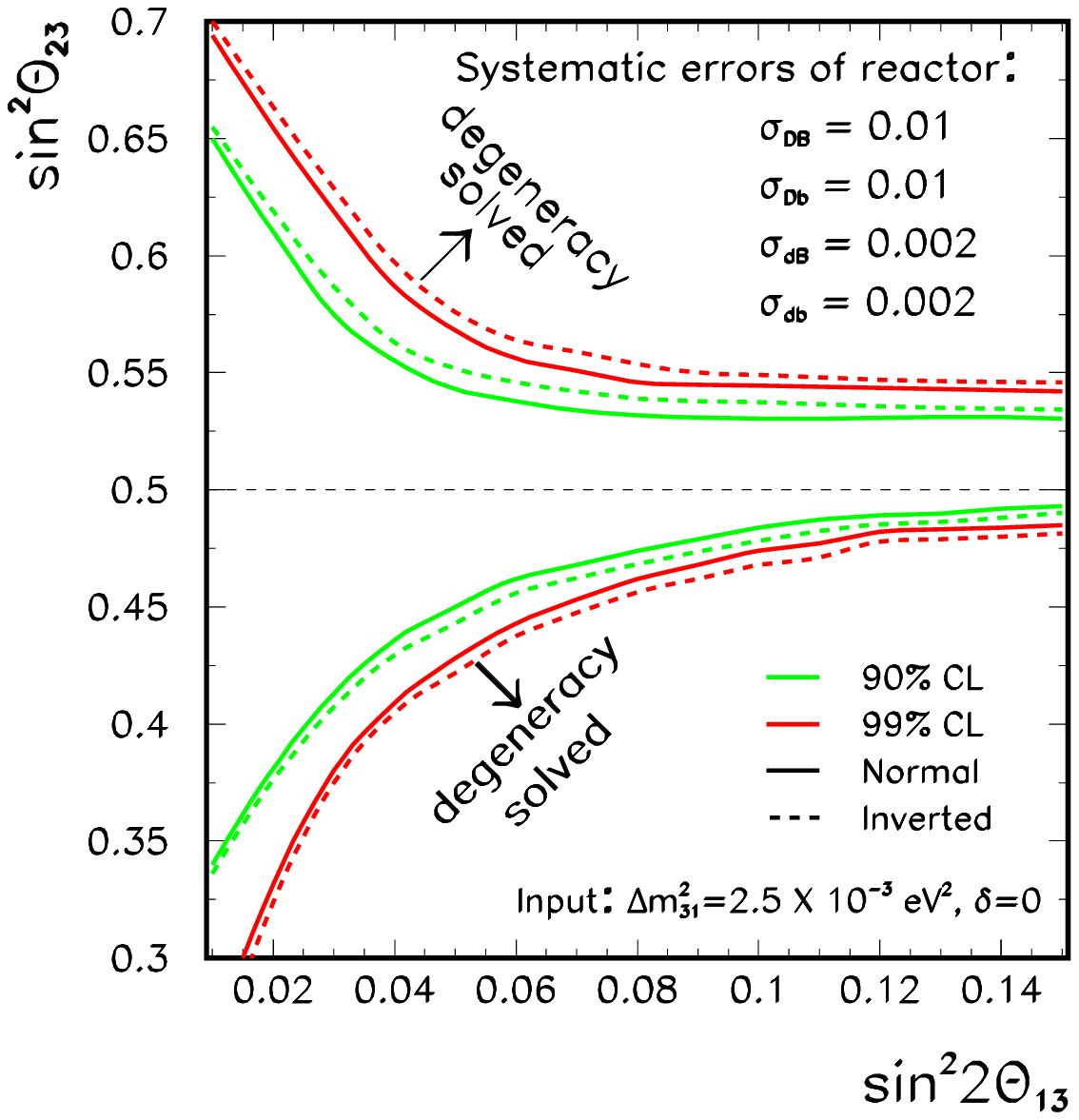}
\end{center}
\vglue -3.4cm
\caption{
The region in $\sin^2 2\theta_{13} - \sin^2 \theta_{23}$ space 
where the $\theta_{23}$ octant degeneracy can be resolved 
at 90\% (thin green) and 99\% (thick red) CL. 
The solid (dashed) curve is for the case of taking the normal (inverted) 
hierarchy to perform the fit, assuming the normal hierarchy as input. 
Conservative systematic errors, as indicated in the figure, are
considered here.
Taken from \cite{resolve23}. 
}
\label{region_resolved}
\end{figure}

Detecting the solar oscillation effect is not the unique way of 
resolving the $\theta_{23}$ octant degeneracy.
The alternative methods proposed include, in addition to the 
already mentioned atmospheric neutrino method, 
the reactor accelerator combined method \cite{MSYIS,resolve23}, and 
the atmospheric accelerator combined method \cite{atm-lbl}.

Here, we explain the reactor-accelerator combined method. 
The principle is again very simple; 
The reactor measurement can pick up one of the solutions 
of $\theta_{13}$ because it is a pure measurement of $\theta_{13}$, 
the possibility first explored in \cite{MSYIS}. 
This principle is explained in Fig.~\ref{23fig1} and in the caption further, 
which are taken from \cite{resolve23}. 
This reference gives a detailed quantitative analysis of the sensitivity 
achievable by the accelerator-reactor combined method.

In Fig.~\ref{region_resolved} presented is the region in 
$\sin^2 2\theta_{13} - \sin^2 \theta_{23}$ space 
where the $\theta_{23}$ octant degeneracy can be resolved. 
The upper and the lower figures in Fig.~\ref{region_resolved} 
are with a relatively pessimistic and an optimistic systematic 
errors, respectively, as indicated in the figures. 
For definition of the errors and details of the analysis procedure, 
see \cite{resolve23}. 
By comparing Fig.~\ref{region_resolved} with Fig.~\ref{sensitivity-theta23-octant}, 
we observe that the sensitivity achievable by the reactor-accelerator 
combined method surpasses that of T2KK in large $\theta_{13}$ region, 
$\sin^2 2\theta_{13} \gsim 0.03 - 0.05$, the critical value very 
dependent of the systematic errors.

\section{How to proceed; Confrontation of theoretical ideas to experiments}
\label{proceed}

In the bottom-up approach to the origin of neutrino mass and 
the mixing it is important to test various phenomenologically 
motivated ideas experimentally. 
In this article we discuss only one example, 
the quark-lepton complementarity (QLC) \cite{QLC}. 
and briefly mention about the $\mu \leftrightarrow \tau$ exchange symmetry. 
An extensive list of the relevant references for the $\mu \leftrightarrow \tau$ 
symmetry, which is too long to quote in this manuscript,  
may be found in \cite{resolve23,moha-smi}.

The empirically suggested relation 
\begin{eqnarray}
\theta_{12} + \theta_C = \frac{\pi}{4}, 
\label{qlc}
\end{eqnarray}
with $\theta_C$ being the Cabibbo angle is under active investigation 
\cite{QLC_rev} and is dubbed as the QLC relation. 
If not accidental, it may suggest a new way of thinking on how quarks 
and leptons are unified.  
It may have extension to the 2-3 sector, 
$\theta_{23}^{(lepton)} + \theta_{23}^{(quark)}  = \frac{\pi}{4}$.

We now discuss how the relation (\ref{qlc}) can be tested 
experimentally. 
Since the Cabibbo angle is measured in an enormous 
precision as emphasized earlier, the real problem is to what accuracy the 
solar angle $\theta_{12}$ can be measured experimentally.
At this moment there exist two approaches to measure $\theta_{12}$ accurately. 
The first one is a natural extension of the method by which 
$\theta_{12}$ is determined today, namely, 
combining the solar and the KamLAND experiments. 
The other one is to create a dedicated new reactor experiment with 
detector at around the first oscillation maximum of reactor neutrino oscillation, ``SADO'' (see below). 
Let me briefly explain about the basic ideas behind them one by one.

\subsection{Solar-KamLAND method}

Combining the solar and the KamLAND experiments is powerful, 
assuming CPT invariance, 
because solar neutrino measurement is good at constraining 
$\theta_{12}$ and KamLAND determines with high precision 
the other parameter $\Delta m^2_{21}$. 
The feature makes the analysis of the solar neutrino parameter 
determination essentially 1-dimensional. 
The former characteristics is particularly clear from the fact that 
the ratio of CC to NC rates in SNO directly measures 
$\sin^2{\theta}_{12}$ in the LMA solution. 
The current data allows accuracy of determination of $\sin^2{\theta}_{12}$ 
of about $\simeq12$\% (2 DOF) (the last reference in \cite{solar}.) 
Further progress in measurement in SNO and KamLAND may 
improve the accuracy by a factor of $\sim$ 2 but not too much beyond that.


\begin{table}
\vglue 0.2cm
\small
\begin{tabular}{c|cc}
\hline
\ Experiments\  & \ $\delta s^2_{12}/s^2_{12}$  at 68.27\% CL \ 
                & \ $\delta s^2_{12}/s^2_{12}$  at 99.73\% CL \  \\
\hline
 Solar+ KL (present)  & $ 8 $ \%  
                      & $ 26  $ \%   \\
\hline
 Solar+ KL (3 yr)  & $ 7 $ \% 
                   & $ 20 $ \%   \\
\hline
 Solar+ KL (3 yr) + pp (1\%) &  $ 4 $ \%  
                 & $ 11$ \%  \\
\hline
\multicolumn{3}{c}{54 km}\\
\hline
 SADO  for 10 $\text{GWth} \cdot$kt$\cdot$yr  &  4.6 \% \   (5.0 \%)  
                 &  12.2 \%  \ (12.9 \%) \\
\hline
 SADO for 20 $\text{GWth}  \cdot$kt$\cdot$yr
                 &  3.4 \% \  (3.8 \%)
                 &  8.8 \% \  (9.5 \%) \\
\hline
 SADO for 60 $\text{GWth}  \cdot$kt$\cdot$yr
                 & 2.1 \%  \   (2.4 \%)
                 & 5.5  \% \   (6.2 \%) \\
\hline
\end{tabular}
\vglue 0.4cm
\caption[aaa]{
Comparisons of fractional errors of the experimentally determined mixing angle, 
$\delta s^2_{12}/s^2_{12} \equiv \delta (\sin^2\theta_{12}) / \sin^2\theta_{12}$, 
by current and future solar neutrino experiments and KamLAND (KL), 
obtained from Tables 3 and 8 of Ref.~\cite{bahcall-pena}, versus that by SADO$_{\text{single}}$, which means to ignore all the other reactors 
than Kashiwazaki-Kariwa, obtained at 68.27\% 
and 99.73\% CL for 1 DOF in \cite{sado1}.
The numbers in parentheses are for SADO$_{\text{multi}}$, which takes 
into account all 16 reactors all over Japan. 
}
\vglue 0.2cm
\label{solar-KL}
\end{table}

\begin{figure}[htbp]
\begin{center}
\vglue 0.4cm
\includegraphics[width=0.64\textwidth]{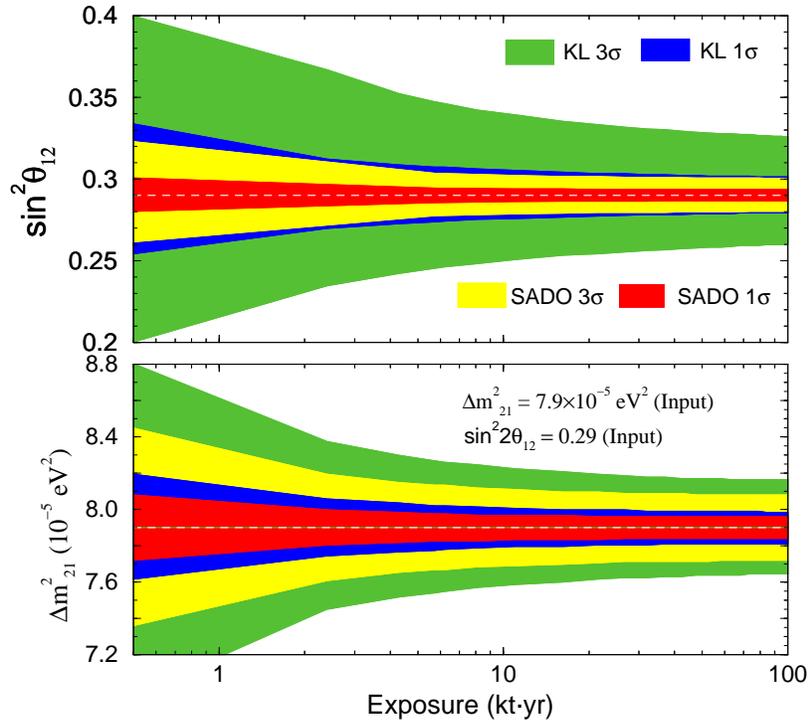}
\end{center}
\vglue -0.2cm
\caption{Accuracies of determination of 
$\sin^2\theta_{12}$ (upper panel) and 
$\Delta m^2_{21}$ (lower panel) reachable by 
KamLAND and SADO (both 1 DOF) are compared with the 
same systematic error of 4\%. 
The geo-neutrino contribution was switched off. 
Taken from \cite{sado2}. 
}
\label{dist-dep}
\end{figure}

If one wants to improve substantially the accuracy of 
$\theta_{12}$ determination, the existing solar neutrino experiments 
are not quite enough. 
Measurement of low-energy pp and $^7$Be neutrinos is 
particularly useful by exploring vacuum oscillation regime. 
Fortunately, varying proposal  for such low energy solar neutrino 
measurement are available in the world \cite{nakahata}. 
Measurement of $^7$Be neutrinos is attempted in Borexino \cite{borexino} 
and in KamLAND \cite{KL-solar}.

The improvement that is made possible by these additional measurement is 
thoroughly discussed by Bahcall and Pe{\~n}a-Garay \cite{bahcall-pena}. 
Since the vacuum oscillation is the dominant mechanism at low energies 
measuring pp neutrino rate gives nothing but 
measurement of $\sin^2{2\theta}_{12}$. 
On the other hand, $^7$Be neutrino may carry unique informations 
of oscillation parameters due to its characteristic feature of 
monochromatic energy. 
The solar-KamLAND method will allow us to determine $\sin^2{\theta}_{12}$ 
to 4\% level at 1$\sigma$ CL \cite{bahcall-pena}. 
In the upper panels of Table~\ref{solar-KL}, 
we tabulate the sensitivities (1 DOF) 
currently obtained and expected by the future measurement. 

\subsection{SADO; Several-tens of km Antineutrino DetectOr}

\begin{figure}[htbp]
\begin{center}
\vglue -2.0cm
\includegraphics[width=0.66\textwidth]{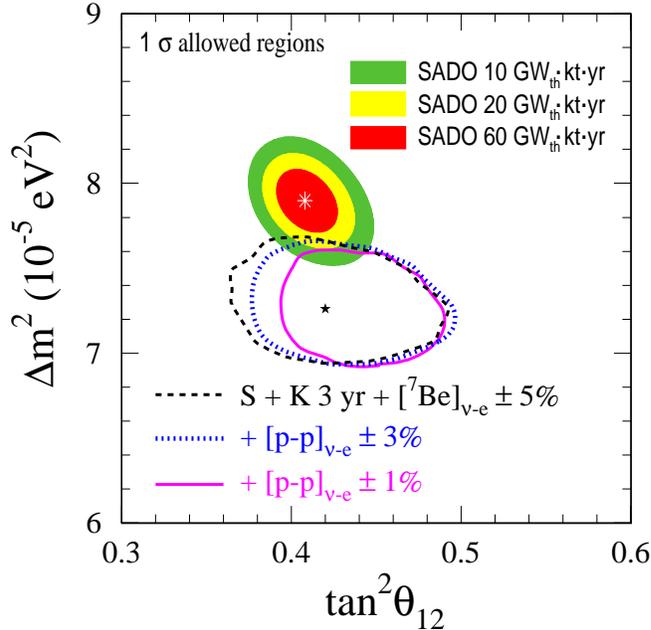}
\end{center}
\vglue -4.8cm
\caption{
SADO's sensitivity contours are plotted in 
$\tan^2\theta_{12}$-$\Delta m^2_{21}$ space and are 
overlaid on Fig.6 of the roadmap paper \cite{bahcall-pena}, 
in which the sensitivities 
of solar-KamLAND combined method are presented. 
The errors are defined both with 2 DOF. 
Taken from \cite{sado2}. 
}
\label{solar-sado}
\end{figure}

Though natural and profitable as a dual-purpose experiment for 
both $\theta_{12}$ and solar flux measurement the solar-KamLAND 
method is not the unique possibility  for reaching the region of the 
highest sensitivity for $\theta_{12}$. 
The most traditional way of measuring mixing angles at the highest 
possible sensitivities is either to tune beam energy to the oscillation 
maximum (for example \cite{T2K} which is for $\sin^2{2\theta}_{23}$), 
or to set up a detector at baseline corresponding to it as employed 
by various reactor experiments to measure $\theta_{13}$ 
\cite{MSYIS,reactor_white}. 
It is also notable that the first proposal of prototype superbeam 
experiment for detecting CP violation \cite{lowECP} entailed 
in a setup at around the first oscillation maximum.

For $\theta_{12}$ the latter method can be applied to reactor neutrinos 
and in fact a concrete idea for possible experimental setup for dedicated 
reactor $\theta_{12}$ is worked out in detail \cite{sado1,sado2}.
See also \cite{reactor12} for the related proposals with reactor neutrinos. 
The type of experiment is dubbed in \cite{sado1} 
as  ``SADO'', an acronym of 
{\it Several-tens of km Antineutrino DetectOr} because of the range of 
baseline distance appropriate for the experiments. 
It is a very feasible experiment because it does not require 
extreme reduction of the systematic error to 1\% level, 
as required in the $\theta_{13}$ measurement mentioned above. 
As is demonstrated in \cite{sado1} reduction of the systematic error to 
4\% level would be sufficient if no energy spectrum cut at 
$E_{prompt} = 2.6$ MeV is performed. 
It should be within reach in view of the current KamLAND error  
of 6.5\% \cite{KL_evidence}. 
The effect of geo-neutrino background, which then has to be worried 
about without spectrum cut, is shown to be tolerable 
even for most conservative choice of geo-neutrino model, 
the Fully Radiogenic model  \cite{sado1}.

The accuracy achievable by the dedicated reactor $\theta_{12}$ 
measurement is quite remarkable. It will reach to  
2\% level at 1$\sigma$ CL (1 DOF) for 60 
GW$_{\text{th}}$$\cdot \text{kt} \cdot \text{yr}$ exposure 
as shown in Table~\ref{solar-KL}. 
With Kashiwazaki-Kariwa nuclear reactor complex, it corresponds 
to about 6 years operation for KamLAND size detector. 
It is notable that possible uncertainties due to the 
surrounding reactors are also modest,  
as one can see in Table~\ref{solar-KL}. 
In Fig.~\ref{solar-sado} we show in the two-dimensional space spanned by 
$\tan^2{\theta}_{12}$ and $\Delta m^2_{21}$ the contours of sensitivities 
achievable by the solar-KamLAND method and by the 
dedicated reactor experiment. 
Notice that the measurement is not yet systematics dominated and 
therefore further improvement of the sensitivity is possible by 
gaining more statistics. 
If SADO can run long enough it can go beyond the solar-KamLAND method.

\subsection{$\mu \leftrightarrow \tau$ symmetry}

The $\mu \leftrightarrow \tau$ exchange symmetry is attractive 
because it predicts $\theta_{13}=0$ and $\theta_{23}= \pi/4$ 
in the symmetry limit. 
For extensive references on this symmetry, see e.g., 
\cite{moha-smi,resolve23,grimusZ2}. 
But, since the symmetry is badly broken 
(note that $m_{\tau} \simeq 20~m_{\mu}$), the predictions 
$\theta_{13}=0$ and $\theta_{23}= \pi/4$ cannot be exact. 
It is important to try to compute deviations from the results obtained 
in the symmetry limit.

Now, the question is how can we pick up the right one out of the 
vast majority of the proposed symmetries? 
One way to proceed is to make clear how symmetry breaking affects 
the predictions. 
For example, it is shown in \cite{grimusZ2} that breaking of the $Z_{2}$ 
symmetry tends to prefer larger deviation from the maximal $\theta_{23}$ 
than vanishing $\theta_{13}$. Such study has to be performed in 
an extensive way including various symmetries.

If one can make definitive prediction in a class of models on which octant 
of $\theta_{23}$ is chosen when the $\mu \leftrightarrow \tau$ symmetry 
is broken, one can test such a class of models by resolving the 
$\theta_{23}$ octant degeneracy. 
We have discussed in Sec.~\ref{resolveD} and in 
Sec.~\ref{reactor-accelerator} the ways of how it can be carried out.

\subsection{Comments on precision measurement of $\Delta m^2$}

We have explained in Sec.~\ref{theta23} how the atmospheric 
$\Delta m^2_{32}$, which may be better characterized as 
$\Delta m^2_{\mu \mu}$ \cite{NPZ}, can be determined. 
For various reasons one may want to improve the accuracy of 
determining $\Delta m^2_{32}$ to a sub-percent level.   
Here, we want to remark that unfortunately there is a serious obstacle 
against it; the problem of absolute energy scale error. 
See Appendix of \cite{MNPZ} for an explicit demonstration of this fact. 
It comes from the limitation of the accuracy of calibrating the absolute energy 
of muons in the case of $\nu_{\mu}$ disappearance measurement. 
The current value for the error in Super-Kamiokande is about 2\% at GeV region 
(second reference in \cite{SKatm}) 
and apparently no concrete idea has been emerged to improve it. 
It is believed to be a limiting factor in $\Delta m^2_{32}$ determination 
in much higher statistics region enabled by T2K II with Hyper-Kamiokande.

We note that there is the unique case which is free from the problem 
of energy scale error; 
the recently proposed resonant $\bar{\nu}_{e}$ absorption reaction 
enhanced by M\"ossbauer effect \cite{raghavan}. 
This method utilizes the recoilless resonant absorption reaction, 
$\bar{\nu}_{e} + ^{3}\mbox{He} + \mbox{orbital e}^{-} \rightarrow \, ^{3}\mbox{H}$, 
with monochromatic $\bar{\nu}_{e}$ beam from the T-conjugate 
bound state beta decay, 
$^{3}\mbox{H} \rightarrow ^{3}\mbox{He} + \mbox{orbital e}^{-} + \bar{\nu}_{e}$.   
When the source atoms are embedded into a solid the energy 
width of the beam is estimated to be $\sim 10^{-11}$ eV, 
which is utterly negligible. 
If feasible experimentally, the monochromatic nature of the beam 
may allow accurate measurement of $\Delta m^2_{31}$ to 
$\simeq(0.3/\sin^2 2\theta_{13})\%$ at 1$\sigma$ CL \cite{mina-uchi}. 
We emphasize that it gives a very rare chance of achieving 
a sub-percent level determination of $\Delta m^2_{31}$.

\section{Conclusion}
\label{conclude}

I have tried to give an overview of neutrino physics emphasizing 
the experimental activities in the near future. 
I must admit that this is a personal overview, not mentioning 
very many important subjects and projects in all over the world, and 
I have to apologize for that. 
But, I tried to give a coherent view which largely come from 
the works I did in the last several years. 
I feel it appropriate to emphasize that we have learned a lot during 
the golden era of neutrino physics. 
But, it seems obvious to me that we have done 
only a half and new surprises are waiting for us in the future.

\begin{acknowledgments}
I would like to thank all of my collaborators, in particular 
Takaaki Kajita, Hiroshi Nunokawa, Masaki Ishitsuka, Shoei Nakayama, 
Renata Zukanovich Funchal, Stephen Parke, Hiroaki Sugiyama, and 
Shoichi Uchinami for fruitful collaborations and useful discussions. 
This work was supported in part by the Grant-in-Aid for Scientific Research, 
No. 16340078, Japan Society for the Promotion of Science. 
\end{acknowledgments}

\end{document}